\documentclass[12pt,a4paper]{article}
\usepackage{latexsym}
\usepackage{amssymb}
\usepackage{epsfig}
\usepackage{float}
\usepackage[T1]{fontenc}
\usepackage[latin1]{inputenc}
\usepackage{graphics}
\usepackage{graphicx}
\usepackage{multicol}
\usepackage{pslatex}
\usepackage{t1enc}
\usepackage{units}
\usepackage{lscape}
\usepackage{slashbox}
\usepackage{amsmath,eepic}
\usepackage{multirow}

\newcommand{\di}{\ensuremath{\mathrm{d}}}

\newcommand{\DMS}{\mbox{\ensuremath{\mathbf{\scriptstyle^-}}\negthickspace\negthickspace {\sl D}}}

\renewcommand{\thefootnote}{\fnsymbol{footnote}}

\newcommand{\0}{$\phantom{0}$}

\begin{document}

\begin{center}
%tittel 

{\bf \large Prediction of Transport Properties by Molecular Simulation: }

{\bf \large Methanol and Ethanol and their mixture}\bigskip

Gabriela Guevara-Carrion\footnotemark[2], Carlos Nieto-Draghi\footnotemark[3], Jadran Vrabec\footnotemark[4]\footnote[1]{To whom correspondence should be addressed, tel.: +49-711/685-66107, fax: +49-711/685-66140, email: vrabec@itt.uni-stuttgart.de}, and Hans Hasse\footnotemark[2]
\end{center}

% Address
\noindent{
{\rm{$\footnotemark[2]$\emph{Laboratory for Engineering Thermodynamics, University Kaiserslautern, 67663 Kaiserslautern, Germany }\\}}}
{\rm{$\footnotemark[3]$\emph{IFP, 1-4 Avenue de Bois Pr\'eau, 92852 Rueil-Malmaison Cedex, France }\\}}
{\rm{$\footnotemark[4]$\emph{Institute of Thermodynamics
and Thermal Process Engineering, University Stuttgart, 70550 Stuttgart, Germany }\\}}

\renewcommand{\thefootnote}{\alph{footnote[1]}}
\baselineskip25pt

\vskip3cm Number of pages: 69

Number of tables: 6

Number of figures: 23

\clearpage

% Abstract
\begin{abstract}
\baselineskip25pt

%{\bf ABSTRACT}\\
\noindent 
Transport properties of liquid methanol and ethanol are predicted by mole\-cular dynamics simulation. The molecular models for the alcohols are rigid, non-polarizable and of united-atom type. They were developed in preceding work using experimental vapor-liquid equilibrium data only. Self- and Maxwell-Stefan diffusion coefficients as well as the shear viscosity of metha\-nol, ethanol and their binary mixture are determined using equilibrium molecular dynamics and the Green-Kubo formalism. Non-equilibrium molecular dynamics is used for predicting the thermal conductivity of the two pure substances. The transport properties of the fluids are calculated over a wide temperature range at ambient pressure and compared with experimental and simulation data from the literature. Overall, a very good agreement with the experiment is found. For instance, the self-diffusion coefficient and the shear viscosity are predicted with average deviations of less 8\% for the pure alcohols and 12\% for the mixture. The predicted thermal conductivity agrees in average within 5\% with the experimental data. Additionally, some velocity and shear viscosity autocorrelation functions are presented and discussed. Radial distribution functions for ethanol are also presented. The predicted excess volume, excess enthalpy and the vapor-liquid equilibrium of the binary mixture methanol + ethanol are assessed and the vapor-liquid equilibrium agree well with experimental data.
\end{abstract}
\bigskip


\clearpage

\section{Introduction}

There has been significant progress in recent years to study and predict both static and dynamic thermophysical properties of fluids by molecular simulation. This is particularly interesting for transport properties of liquids, where, due to the complexity of the involved physical mechanisms, other theoretical approaches are often unsatisfactory. Hence, molecular dynamics is a promising method for predicting transport properties of liquids. However, the number of publications regarding transport properties for technically relevant liquid systems that achieve quantitatively accurate results is still low. The majority of these publications deals with the self-diffusion coefficient, leaving aside the more demanding Maxwell-Stefan diffusion coefficient, shear viscosity and thermal conductivity. Furthermore, molecular simulation studies on transport properties for complex liquids, e.g. mixtures containing highly polar hydrogen-bonding molecules, are rare. This is also the case for methanol, ethanol and their binary mixture. It should be pointed out that there is a series of articles on transport properties of pure liquid ethanol by Petravi\'{c} et al. \cite{petravic2003, petravic2005v, petravic2005t}. There are several works on transport properties of associating mixtures: Typically, aqueous solutions of methanol \cite{wheeler1998, lucassen1999, hawlicka2000, nieto2003t, wensink2003} or ethanol \cite{nieto2003t, wensink2003, plathe1996, zhang2006, noskov2005} have been studied using molecular dynamics simulation. However, to our knowledge the mixture methanol + ethanol has not yet been regarded in that sense.

Two fundamentally different methods for calculating transport properties by molecular dynamics simulation are available. The equilibrium methods (EMD), using either the Green-Kubo formalism or the Einstein relations, analyze the time dependent response of a fluid system to spontaneous fluctuations. With non-equilibrium molecular dynamics (NEMD), on the other hand, the system's response to an externally applied perturbation is analyzed. NEMD was developed in order to increase the signal to noise ratio and to improve statistics and convergence. Both methods have different advantages and disadvantages, but are comparable in efficiency, as shown e.g. by Dysthe et al. \cite{dysthe1999}.

The success of molecular dynamics simulation to predict thermodynamic properties is highly determined by the potential model used to describe the intermolecular interactions. Numerous molecular models for methanol and ethanol have been proposed in the literature \cite{plathe1996, jorgensen1986, haughney1987, mayo1990, leeuwen1995, gao1995, caldwell1995, jorgensen1996, potoff1999, stubbs2001, chen2001, khare2004, dang2003, shihao2007}. However, the ability of these models for predicting transport properties has been probed predominantly regarding the self-diffusion coefficient, cf. Table \ref{literature}. 

For methanol, mainly the rigid OPLS molecular model from Jorgensen \cite{jorgensen1986} and its modifications proposed by Haughney et al. \cite{haughney1987} as well as by van Leeuwen and Smit \cite{leeuwen1995}, have been used to predict the self-diffusion coefficient. It was shown by comparison of the models by Jorgensen \cite{jorgensen1986} and by Haughney et al. \cite{haughney1987} that the rigid OPLS model and its modifications perform well for the self-diffusion coefficient, deviations to experimental data range from 7 to 20\% \cite{haughney1987}. Van de Ven-Lucassen et al. \cite{lucassen1999} showed that the model from van Leeuwen and Smit \cite{leeuwen1995} predicts the self-diffusion coefficient of methanol at ambient temperature more accurately than the polarizable model of Caldwell and Kollman \cite{caldwell1995}, and than the models by Haughney et al. \cite{haughney1987}. Wheeler et al. \cite{wheeler1997} predicted the shear viscosity of methanol within 10\% of experimental values. The thermal conductivity of methanol at ambient conditions was reported by Nieto-Draghi \cite{nieto2003t} with a deviation of $+11\%$.

Different rigid \cite{jorgensen1996, taylor2003}, semi-flexible \cite{saiz1997}, flexible \cite{shihao2007} and polarizable \cite{noskov2005, sandeep2005} molecular models for ethanol have been used to predict the self-diffusion coefficient. Thereby, it has been found that rigid, united-atom molecular models predict the self-diffusion coefficient more reliably than polarizable, flexible models. Most ethanol models assessed by Wang and Cann \cite{shihao2007} as well as the Jorgensen model \cite{jorgensen1986} investigated by Saiz et al. \cite{saiz1997} overestimate the self-diffusion coefficient rather significantly with deviations ranging from +8 to +50\%. Allowing a potential model for bending results in a further increase in the predicted values \cite{petravic2003}. Gonz\'alez et al. \cite{gonzalez1999} found that rigid, non-polarizable models perform better than the polarizable ones for predictions of the self-diffusion coefficient. The shear viscosity of ethanol has been predicted using EMD \cite{petravic2005v} and NEMD \cite{zhao2007}. Both works achieved good results (average deviations are 11 and 7\%, respectively) using the united-atom model and the all-atom model of Jorgensen et al. \cite{jorgensen1986, jorgensen1996}, respectively. Simulations performed by Petravi\'{c} \cite{petravic2005t} underestimate the thermal conductivity of liquid ethanol near ambient temperatures by approximately $-10$\%.

The present article assesses the predictive power of rigid, non-polarizable, united-atom type Lenard-Jones based models with superimposed point charges for methanol and ethanol developed in prior work of our group \cite{schnabel2007, schnabel2005}. These models have been optimized on the basis of experimental data of vapor pressure and saturated liquid density in the whole temperature range from the triple point to the critical point. No experimental data for transport properties was thereby taken into account. The most significant transport properties, i.e. self-diffusion coefficient, shear viscosity and thermal conductivity, were regarded here for the pure liquids using EMD and NEMD. Furthermore, the two molecular models were tested regarding the prediction of self- and Maxwell-Stefan diffusion coefficients and shear viscosity of the binary mixture. Additionally, binary vapor-liquid equilibria as well as excess volume and excess enthalpy were predicted. The goal of this study is to demonstrate the ability of rigid, non-polarizable molecular models, adjusted to pure substance vapor-liquid equilibria only,to accurately describe the transport properties of strongly polar and H-bonding liquids.

This work is organized as follows: Firstly, the employed molecular models are briefly described. Then, the simulation techniques are explained, the results for self-diffusion coefficient, shear viscosity and thermal conductivity of the pure liquids are presented and compared to experimental and simulation data from the literature. The self- and Maxwell-Stefan diffusion coefficients as well as the shear viscosity for the mixture methanol + ethanol are presented and compared with experimental data where possible. Furthermore, the velocity and shear viscosity autocorrelation functions are described and discussed. Simulated radial distribution functions of ethanol are also presented. Thereafter, the simulation results for vapor-liquid equilibria, excess volume and excess enthalpy for the binary mixture are given in comparison to experimental values. Finally, conclusions are drawn. Simulation details are summarized in the Appendix.

\section{Molecular models} 

%\subsection*{Potential Models}
Throughout this work, rigid, non-polarizable molecular models of united-atom type from earlier work of our group \cite{schnabel2007, schnabel2005} were used. Both models account for the intermolecular interactions, including hydrogen bonding, by a set of united-atom Lennard-Jones (LJ) sites and superimposed point charges. Hence, the potential energy $u_{ij}$ between two alcohol molecules $i$ and $j$ can be written as

\begin{equation}
	u_{ij}\left( r_{ijab}\right) =\sum_{a=1}^{n}\sum_{b=1}^{m}4\epsilon_{ab}\left[ \left( \frac{\sigma_{ab}}{r_{ijab}}\right)^{12}-\left( \frac{\sigma_{ab}}{r_{ijab}}\right)^6\right] + \frac{q_{ia}q_{jb}}{4\pi\varepsilon_{0}r_{ijab}}\, , \label{equ_pot}
\end {equation}

\noindent where $a$ is the site index of molecule $i$, $b$ the site index of molecule $j$, and $n$ and $m$ are the numbers of interaction sites of molecules $i$ and $j$, respectively. $r_{ijab}$ represents the site-site distance between molecules $i$ and $j$. The LJ size and energy parameters are $\sigma_{ab}$ and $\epsilon_{ab}$. $q_{ia}$ and $q_{jb}$ are the point charges located at the sites $a$ and $b$ on the molecules $i$ and $j$, and $\varepsilon_{0}$ is the permittivity of vacuum. 

The methanol model has two LJ sites, i.e. one for the methyl group and one for the hydroxyl group. Additionally, three point charges are superimposed, i.e. one at each of the LJ site centers and one at the nucleus position of the hydroxyl hydrogen. Ethanol was modeled by three LJ sites, i.e. one for the methyl, the methylene and the hydroxyl group each. Three point charges are located here as well on the methylene and hydroxyl LJ site centers and on the nucleus position of the hydroxyl hydrogen. 

The parameter set of the methanol model was obtained using the optimization procedure of Stoll \cite{stoll2005} on the basis of the geometry of the model from van Leeuwen and Smit \cite{leeuwen1995}. The geometry of the ethanol model was derived from ab initio quantum chemistry calculations. The AUA4 parameters of Ungerer et al. \cite{ungerer2000} were applied for the methyl and methylene LJ sites. The point charge, LJ size and energy parameters as well as the offset of the hydroxyl LJ site were fitted to experimental vapor pressure and saturated liquid density. It should be pointed out that no information on transport or mixture properties was included in the optimization procedures of any molecular model employed here so that the data presented below is strictly predictive. The model parameters were taken from \cite{schnabel2007, schnabel2005} and are summarized in Table \ref{tab_meoh_pot}.

To define a molecular model for a binary mixture on the basis of pairwise additive pure substance models, only the unlike interactions have to be specified. In case of polar interaction sites, i.e. point charges here, this can straightforwardly be done using the laws of electrostatics. However, for the unlike LJ parameters there is no physically sound approach \cite{haslam2008}, and for predictions combination rules have to be employed. Here, the interaction between unlike LJ sites of two molecules are calculated applying the Lorentz-Berthelot combining rules:

\begin{equation}\label{LBCR1}
\sigma_{ab} = \frac{\sigma_{aa}+\sigma_{bb}}{2}\,,
\end{equation}

\noindent and

\begin{equation}
\epsilon_{ab} = \sqrt{\epsilon_{aa}\epsilon_{bb}}\,.\label{LBCR2}
\end{equation}

\section{Methodology \label{metho}}

Dynamic properties can be obtained from EMD simulations by means of the Green-Kubo formalism \cite{green1954,kubo1957}. These equations are a direct relationship between a transport coefficient and the time integral of an autocorrelation function of a particular microscopic flux in a system in equilibrium. This method was used in this work to calculate self- and Maxwell-Stefan diffusion coefficients as well as the shear viscosity. The thermal conductivity was calculated using a modified version of the boundary driven (BD) reverse NEMD algorithm from M\"uller-Plathe \cite{plathe1997}, also referred to as PeX algorithm. 

\subsection {Diffusion coefficients}

The self-diffusion coefficient $D_{i}$ is related to the mass flux of single molecules within a fluid. Therefore, the relevant Green-Kubo expression is based on the individual molecule velocity autocorrelation function as follows

\begin{equation}
D_{i}=\frac{1}{3N}\int_0^{\infty}\di t~\big\langle\mathbf{v}_k(t)\cdot\mathbf{v}_k(0)\big\rangle,\label{self}
\end{equation}

\noindent where $\mathbf{v}_k(t)$ is the center of mass velocity vector of molecule $k$ at some time $t$, and $<$...$>$ denotes the ensemble average. Eq. (\ref{self}) is an average over all $N$ molecules in a simulation, since all contribute to the self-diffusion coefficient. In a binary mixture, the Maxwell-Stefan diffusion coefficient $\DMS_{ij}$ can also be written in terms of the center of mass velocity

\begin{equation}\label{MS}
\DMS_{ij}=\frac{x_j}{3N_i}\left(\frac{x_i M_i+x_j M_j}{x_jM_j}\right)^2\int_0^{\infty}\di t~\big\langle\sum_{k=1}^{N_i} \mathbf{v}_i^k(t)\cdot\sum_{k=1}^{N_i} \mathbf{v}_i^k(0)\big\rangle,
\end{equation}

\noindent where $M_i$ and $x_i$ are the molar mass and mole fraction of species $i$. From the expression above, the collective character of the Maxwell-Stefan diffusion coefficient is evident.

Since the present simulations provide self- as well as the Maxwell-Stefan diffusion coefficients simultaneously, a comparison to the simple predictive approach suggested by Darken \cite{darken1948} is possible. The Maxwell-Stefan diffusion coefficient $\DMS_{ij}$ is estimated from the self-diffusion coefficients of the two components $D_{i}$ and $D_{j}$ in the binary mixture \cite{darken1948}

\bigskip
\begin{equation}
\label{darken}
\DMS_{ij}= x_i \cdot D_{i} + x_j\cdot D_{j} \cdot
\end{equation}

This approach was found to be superior to the correlations by Vignes \cite{vignes1966} and by Caldwell and Babb \cite{caldwell1956} in prior work on three binary mixtures \cite{fernandez2005}. However, it is of little practical use due to the fact that it relies on the self-diffusion coefficients in the mixture, which are rarely available.

%Shear viscosity
\subsection {Shear viscosity}
The shear viscosity $\eta$, as defined by Newton's "law" of viscosity, is a measure of the resistance of a fluid
to a shearing force \cite{hoheisel1994}. It is associated with momentum transport under the influence of velocity gradients. Hence, the shear viscosity can be related to the time autocorrelation function of the off-diagonal elements of the stress tensor $J_p^{xy}$ \cite{gubbins1972}

\begin{equation}\label{shear}
\eta=\frac{1}{Vk_BT}\int_{0}^{\infty} \di t~\big\langle J^{xy}_p(t)\cdot J^{xy}_p(0)\big\rangle,
\end{equation}

\noindent where $V$ stands for the molar volume, $k_B$ is the Boltzmann constant, and $T$ denotes the temperature. Averaging over all three independent elements of the stress tensor, i.e. $J_p^{xy}$, $J_p^{xz}$,and $J_p^{yz}$, improves the statistics of the simulation. The component $J_p^{xy}$ of the microscopic stress tensor $\textbf{J}_p$ is given in terms of the molecule positions and velocities by \cite{hoheisel1994}

\bigskip
\begin{equation}
J_p^{xy}=\sum_{i=1}^N m v_{i}^x v_{i}^y -\frac{1}{2}\sum_{i=1} ^N \sum_{j \neq i}^N \sum_{k=1}^n
\sum_{l=1}^n r_{i j}^x \frac{\partial u_{ij}}{\partial r_{k l}^y}\label{shear_t}.
\end{equation}

\noindent Here, the lower indices $l$ and $k$ count the interaction sites, and the upper indices $x$ and $y$ denote the spatial vector components, e.g. for velocity $v_i^x$ or site-site distance $r_{ij}^x$. The first term of Eq. (\ref{shear_t}) is the kinetic energy contribution and the second term is the potential energy contribution to the shear viscosity. Consequently, the Green-Kubo integral (\ref{shear}) can be decomposed into three parts, i.e. the kinetic energy contribution, the potential energy contribution and the mixed kinetic-potential energy contribution \cite{hoheisel1994}. Eqs. (\ref{shear}) and (\ref{shear_t}) may directly be applied to mixtures.

%Thermal conductivity
\subsection {Thermal conductivity}
The thermal conductivity $\lambda$, as defined by Fourier's "law" of heat conduction, characterizes the capability of a substance for molecular energy transport driven by temperature gradients. In BD-NEMD simulation used in this work, two slabs of a given thickness, large enough to contain a sufficiently large number of molecules on average but much smaller than the total length of the simulation volume in $z$ direction, are defined. In order to create a heat flux in $z$ direction, the molecule with the highest kinetic energy in the cold slab is selected with a frequency $\upsilon$ to perform a hypothetical elastic collision with the molecule having the lowest kinetic energy in the hot slab \cite{nieto2003}. After such a virtual collision the two molecules have exchanged their momentum without changing their respective position in space. The new velocity of the molecule in the cold region is then

\bigskip
\begin{equation}\label{velcold}
v^{new}_{c}= -v^{old}_{c}+ \frac{m_{c}v^{old}_{c}+ m_{h}v^{old}_{h}}{m_{c}+m_{h}},
\end{equation}

\noindent and, for the molecule in the hot region

\bigskip
\begin{equation}\label{velhot}
v^{new}_{h}= -v^{old}_{h}+ \frac{m_{c}v^{old}_{c}+ m_{h}v^{old}_{h}}{m_{c}+m_{h}},
\end{equation}

\noindent where $m_{c}$ and $m_{h}$ are the respective masses of the selected molecules in the cold and hot slabs. $v^{old}_{h}$ and $v^{new}_{h}$ stand, respectively, for the velocities of the molecules before and after the collision. 

During the elastic collision the molecules exchange their momentum. Thus, the total momentum and the total energy of the system remain constant. The kinetic energy exchanged by means of these elastic collisions determines the energy flux in the system. The heat flow density in the steady state is then given by 

\bigskip
\begin{equation}\label{momexch}
\left\langle J_{c}\right\rangle_{t}=\frac{1}{2At}\sum_{\rm transfers} \frac{m_{h}}{2}\left((v^{new}_{h})^{2} - (v^{old}_{h})^{2}\right).
\end{equation}

\noindent In this expression, $A$ is the cross sectional area in $x$ and $y$ direction, $t$ is the time interval of measure in the steady state during which a given number of momentum transfers have occurred. $\left\langle J_{c}\right\rangle_{t}$ is the heat flux in $z$ direction, which can indistinctly be evaluated in the cold or the hot slab. It is required that the number of molecules in both slabs is sufficiently large so that the intermolecular interactions within the slabs will properly thermalize the velocity distribution before a new collision takes place. The thermal gradient $\nabla T_{z}$ is obtained from a linear fit of the temperature profiles from the simulation. In the steady state, the thermal conductivity can be obtained by

\bigskip
\begin{equation}\label{thestat}
\lambda= - \frac{\left\langle J_{c}\right\rangle_{t}}{\nabla T_{z}}\cdot
\end{equation}

%Results

\section {Simulation results}

\subsection{Diffusion coefficients}

\begin{figure}[ht]
\begin{center}
\caption[Temperature dependence of the self-diffusion coefficient of pure methanol at $0.1$ MPa. Present simulation results ($\bullet$) are compared to experimental data \cite{karger1990, asahi1998, johnson21956, dullien1972, hurle1985} ({$+$}) and to other simulation results \cite{lucassen1999} ($\triangle$), \cite{haughney1987} ($\Diamond$), \cite{asahi1998} ({\large$\circ$}), and \cite{pereira2001} ($\Box$). Present simulation error bars are within symbol size.]{} \label{figselfmeoh}

\includegraphics[scale = 0.5]{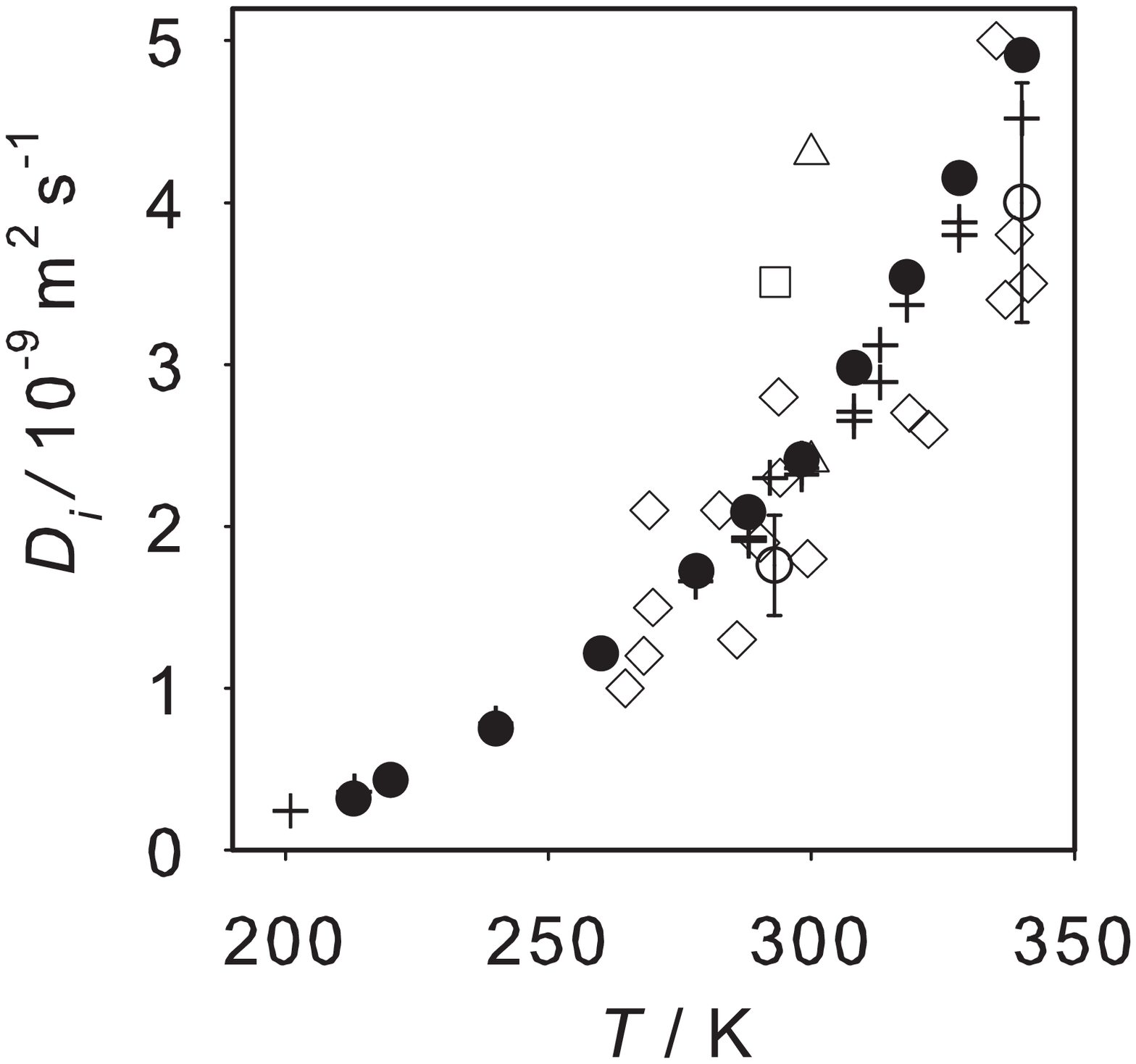}
\end{center}
\end{figure}

\begin{figure}[ht]
\begin{center}
\caption[Temperature dependence of the self-diffusion coefficient of pure ethanol at $0.1$ MPa. Present simulation results ($\bullet$) are compared to experimental data \cite{karger1990, johnson21956, dullien1972, hurle1985, meckl1988} ({$+$}) and to other simulation results \cite{petravic2003} ($\Diamond$), \cite{shihao2007} ({\large$\circ$}), \cite{taylor2003} ($\triangle$), \cite{saiz1997} (star), \cite{sandeep2005} ($\triangledown$), \cite{gonzalez1999} (hexagon), and \cite{pereira2001} ($\Box$). Present simulation error bars are within symbol size.]{} \label{figselfetoh}
\includegraphics[scale = 0.5]{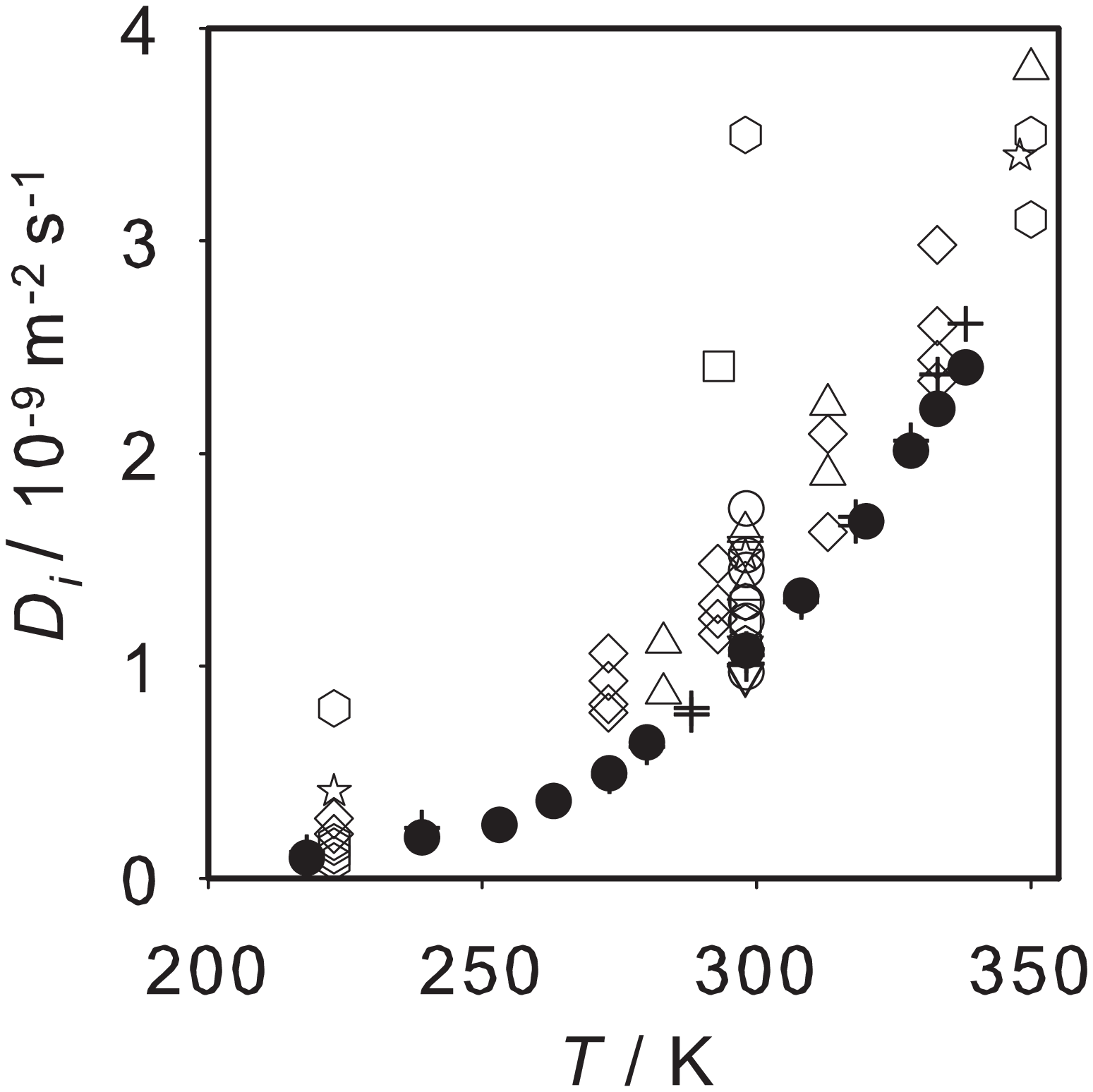}
\end{center}
\end{figure}

The self-diffusion coefficient of pure methanol and pure ethanol was predicted at ambient pressure in the temperature range from $210$ to $340$ K. Present numerical data are given in Table \ref{resmeoh}. Figures \ref{figselfmeoh} and \ref{figselfetoh} show the self-diffusion coefficient for methanol and ethanol as a function of temperature, respectively. The statistical error of the simulation data is estimated to be in the order of $1\%$.

The self-diffusion coefficient of methanol shows a very good agreement with the experimental values. At temperatures below 280 K, the self-diffusion coefficient was underpredicted by approximately $-$7\%. Above ambient temperature, the mobility of the model molecules increases slightly more rapidly than that of real methanol molecules. Hence, for temperatures above 300 K the self-diffusion coefficient was overestimated on average by $+$8\%. However, the significant scatter of the experimental data should be taken into account. In comparison to previous simulation studies from the literature, the present results cover the widest temperature range. The broadest available study by Haughney et al. \cite{haughney1987}, using a modified Jorgensen model, is comparable with present results near ambient temperature where average deviations are 9 \%. However, the data from Haughney et al. is scattering and the self-diffusion coefficient is significantly underestimated above 300 K.

There is an excellent agreement between experimental data and present self-diffusion coefficient for ethanol. The predicted data also correctly reproduces the temperature dependence over the whole temperature range from $239$ to $338$ K. Contrary to previous simulation reports \cite{petravic2003, shihao2007, taylor2003, saiz1997, sandeep2005, gonzalez1999, pereira2001}, where the self-diffusion coefficient of ethanol has mostly been significantly overestimated, deviations of the present data are negative and on average only $-6$\%. This is explained by the different potential model used in this work.

%\subsubsection*{Methanol-Ethanol Mixture}

\begin{figure}[htb]
\begin{center}
\caption[Composition dependence of the self-diffusion coefficients for the mixture methanol + ethanol at $298.15$ K and $0.1$ MPa. Present simulation results for methanol ($\bullet$) and ethanol ($\blacktriangle$) are compared to experimental data \cite{johnson1956} ({\large$\circ$}) and ($\triangle$).]{} \label{figselfmix}
%\vskip1cm
\includegraphics[scale = 0.5]{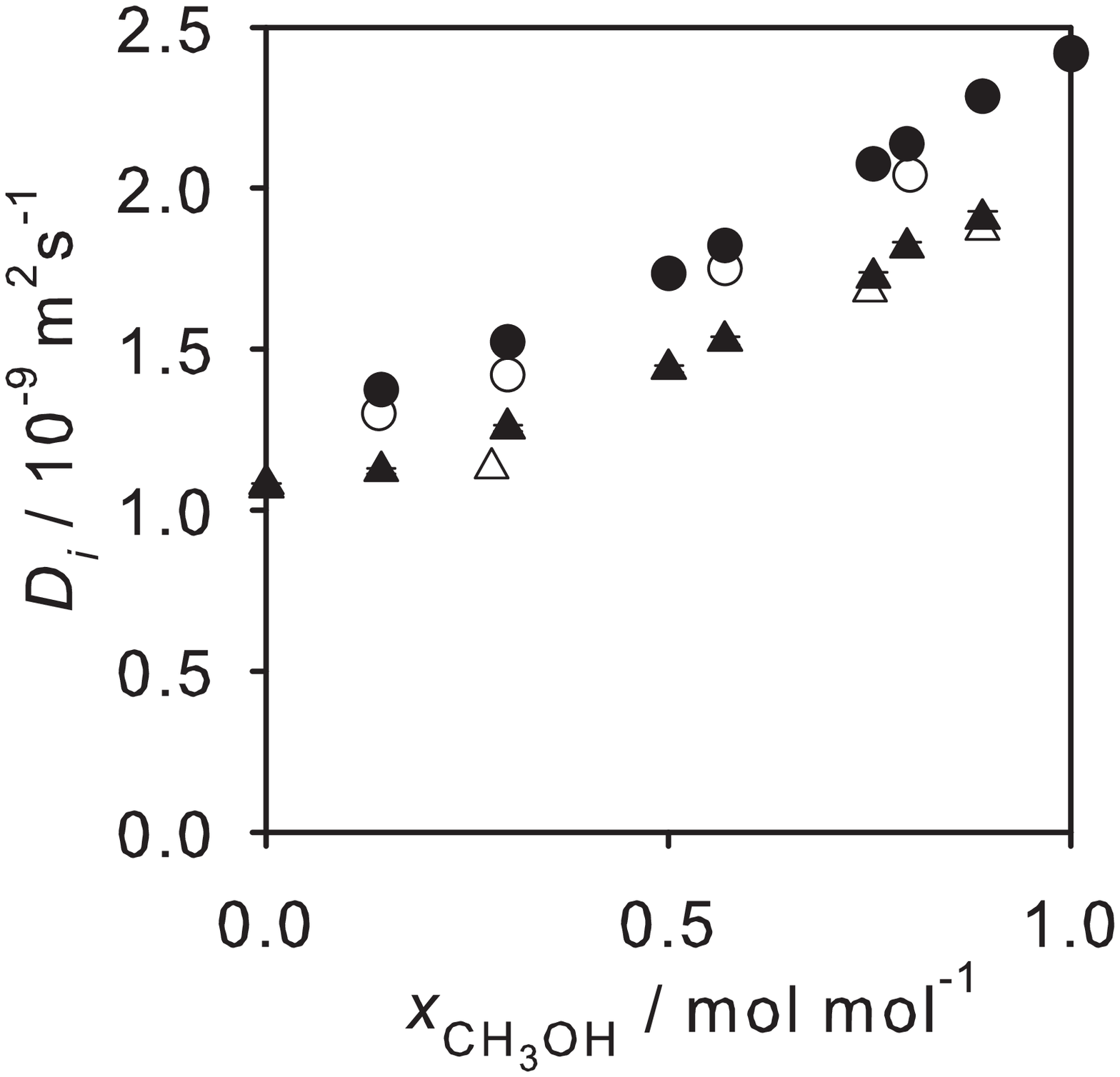}

\end{center}
\end{figure}

Self-diffusion coefficients of methanol and ethanol in their binary mixture were predicted at ambient conditions for the entire composition range. Numerical results are presented in Table \ref{resmix} and plotted in Figure \ref{figselfmix}. The estimated statistical uncertainty of the simulation data is also $1\%$. As shown in Figure \ref{figselfmix}, there is an excellent agreement between predicted self-diffusion coefficients and the experimental data for both alcohols. Present simulation results slightly overestimate the self-diffusion of both alcohols by $+5\%$ on average. Furthermore, the predicted data reproduce the composition dependence well.

\begin{figure}[htb]
\begin{center}
\caption[Composition dependence of the Maxwell-Stefan diffusion coefficient for the mixture methanol + ethanol at $298.15$ K and $0.1$ MPa. Present simulation results ($\bullet$) are compared to Darken's model \cite{darken1948} ({\large$\circ$}), cf. Eq. (\ref{darken})]{} \label{figMS}
%\vskip1cm
\includegraphics[scale = 0.5]{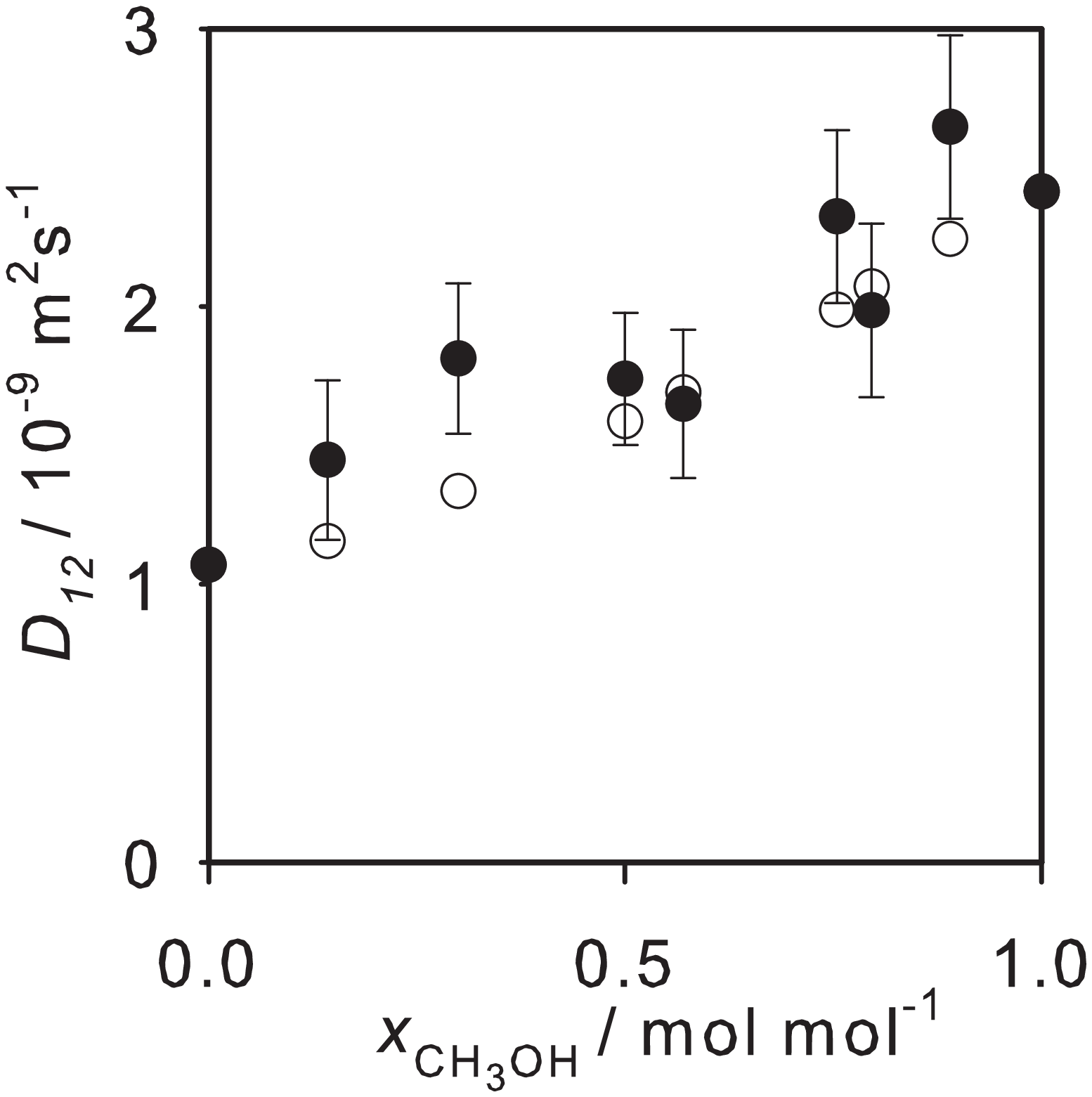}

\end{center}
\end{figure}

The Maxwell-Stefan diffusion coefficient of the mixture methanol + ethanol calculated by molecular simulation is reported in Table \ref{resmix}. In Figure \ref{figMS}, the predicted values are plotted as a function of the composition. The statistical uncertainties are on average $15\%$. These large error bars are attributed to the collective nature of the Maxwell-Stefan diffusion coefficient. Due to the lack of experimental data, present simulation results are only compared to the predictions of Darken's model \cite{darken1948}, cf. Eq. (\ref{darken}). This model can be considered to be a good approximation here, in agreement to previous findings for three other mixtures \cite{fernandez2005}.

\subsection{Shear viscosity}

\begin{figure}[htb]
\begin{center}
\caption[Temperature dependence of the shear viscosity of pure methanol at $0.1$ MPa. Present simulation results using reaction field ($\bullet$) and Edwald sumation ($\blacktriangle$) are compared to experimental data \cite{rauf1983, vargaftik} ($+$) as well as to other EMD ($\Diamond$) and NEMD ({\large$\circ$}) simulation results \cite{wheeler1997}.]{} \label{figviscomeoh}
%\vskip1cm
\includegraphics[scale = 0.5]{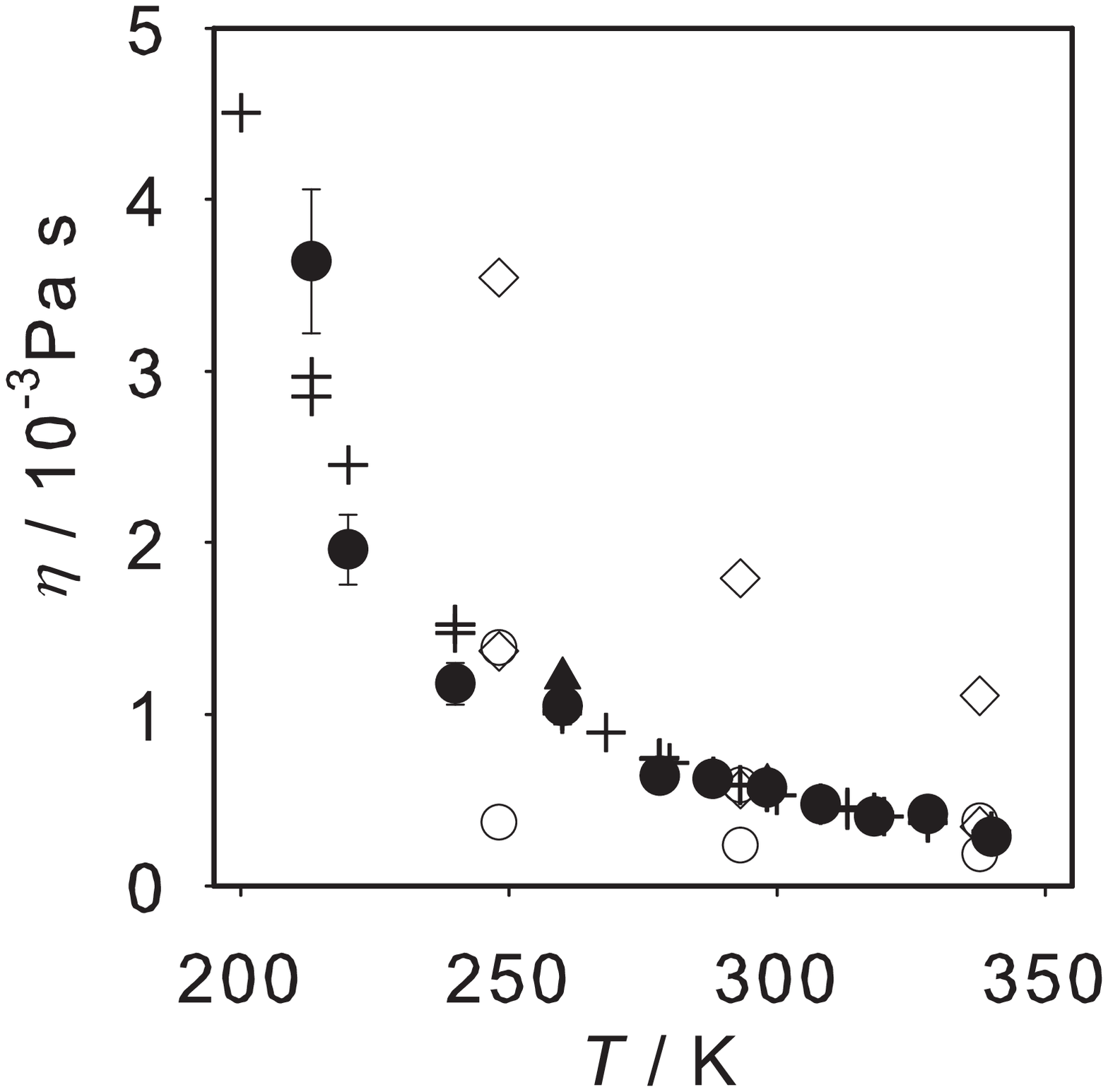}
\end{center}
\end{figure}

\begin{figure}[htb]
\begin{center}
\caption[Temperature dependence of the shear viscosity of pure ethanol at $0.1$ MPa. Present simulation results using reaction field ($\bullet$) and Edwald summation ($\blacktriangle$) are compared to experimental data \cite{rauf1983} ($+$) and to other simulation results \cite{petravic2005v} ($\Diamond$) and \cite{zhao2007} ({\large$\circ$}).]{} \label{figviscoetoh}
\vskip1cm
\includegraphics[scale = 0.5]{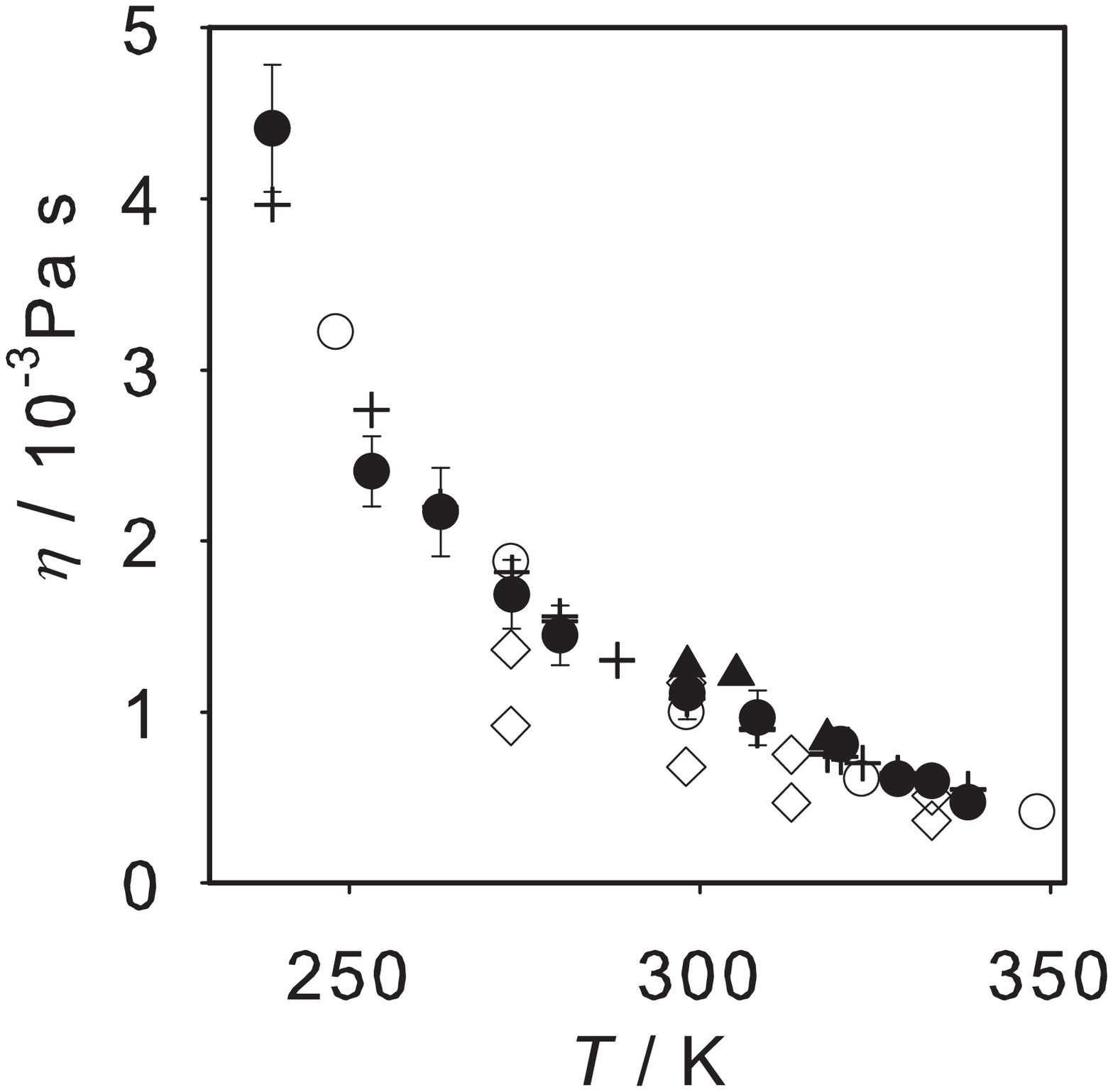}
\end{center}
\end{figure}

The shear viscosity of pure methanol and pure ethanol was predicted at ambient pressure for a temperature range of about 100 K around ambient conditions and is reported in Table \ref{resmeoh}. The temperature dependence of the shear viscosity is shown in Figures \ref{figviscomeoh} and \ref{figviscoetoh} for methanol and ethanol, respectively. Statistical errors are in the range from $10$ to $17\%$.

In the case of methanol, the predicted shear viscosity agrees very well with the experimental data over the whole regarded temperature range with an average deviation of 8\%. At higher temperatures, the predicted viscosities are closer to the experimental ones than at lower temperatures, cf. Figure \ref{figviscomeoh}. It should be pointed out that the highest deviation, being 20\%, was found at the lowest regarded temperature 220 K. Here, the high density in combination with strongly interacting molecules causes large simulation uncertainties due to the long-time behavior of the shear viscosity autocorrelation function. Present shear viscosity predictions are better than the EMD data of Wheeler et al. \cite{wheeler1997} and are comparable with the NEMD Edwald data of the same author. 

The calculated shear viscosity for ethanol shows a good agreement to experimental data with an average deviation of 8\%. It can be noticed in Figure \ref{figviscoetoh} that there is a tendency to underestimate the shear viscosity. Petravi\'{c} and Delhommelle \cite{petravic2005v} also underestimated the shear viscosity using the OPLS model from Jorgensen \cite{jorgensen1986}. At high densities, however, present data better predict the shear viscosity. Zhao et al. \cite{zhao2007} recently reported shear viscosity data using the OPLS-AA model \cite{jorgensen1996} and NEMD which is in very good agreement with experimental data.

The three contributions to the shear viscosity were also calculated. In agreement with the results of Meier et al. \cite{meier2004v}, the potential energy contribution is dominant for both studied alcohols, $\eta_{pp}> 90\%$. In general, for the studied temperature range, the kinetic energy term has a negligible positive contribution to the shear viscosity, $\eta_{kk}< 2\%$. The mixed kinetic-potential energy term may contribute positively or negatively to the shear viscosity. Here, this contribution varies between 0.1 and 8\%.

\begin{figure}[htb]
\begin{center}
\caption[Composition dependence of the shear viscosity for the mixture methanol + ethanol at $298.15$ K and $0.1$ MPa. Present simulation results ($\bullet$) are compared to experimental data \cite{mussche1975} ($+$).]{} \label{figviscomix}
%\vskip1cm
\includegraphics[scale = 0.5]{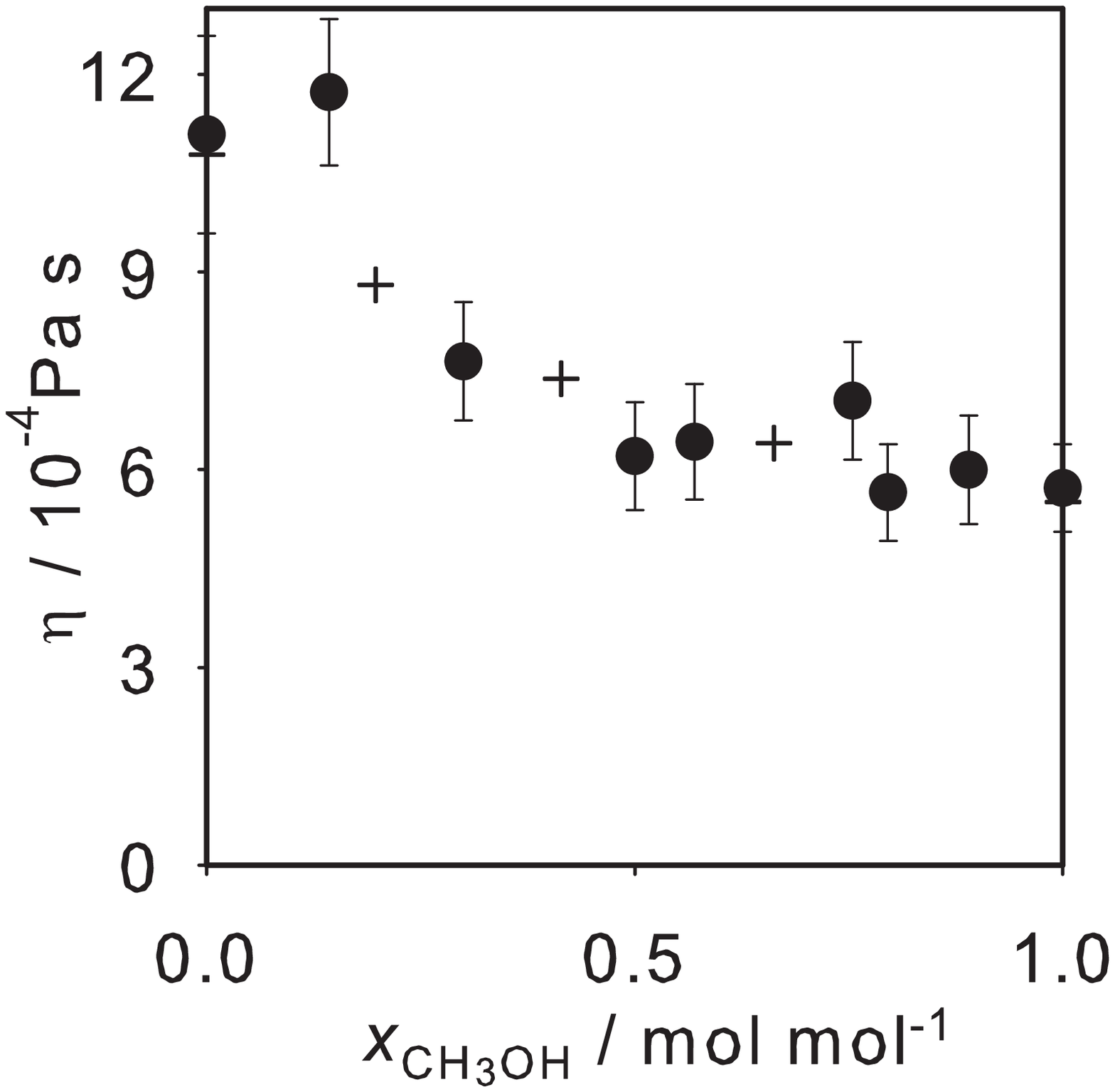}
\end{center}
\end{figure}

The simulated shear viscosity data of the mixture methanol + ethanol is given in Table \ref{resmix} and plotted in Figure \ref{figviscomix} together with the available experimental data. There is a very good agreement between experimental and predicted shear viscosity, being mostly within the simulation uncertainty of around 12\%. Deviations from the experimental data are on average 12\% as well.

\subsection{Thermal conductivity}

\begin{figure}[htb]
\begin{center}
\caption[Temperature dependence of the thermal conductivity of pure methanol at $0.1$ MPa. Present simulation results ($\bullet$) are compared to experimental data \cite{vargaftik, poling, touloukian, CRC} ($+$) and to other simulation results \cite{nieto2003t} ({\large$\circ$}).]{} \label{figtcmeoh}
%\vskip1cm
\includegraphics[scale = 0.5]{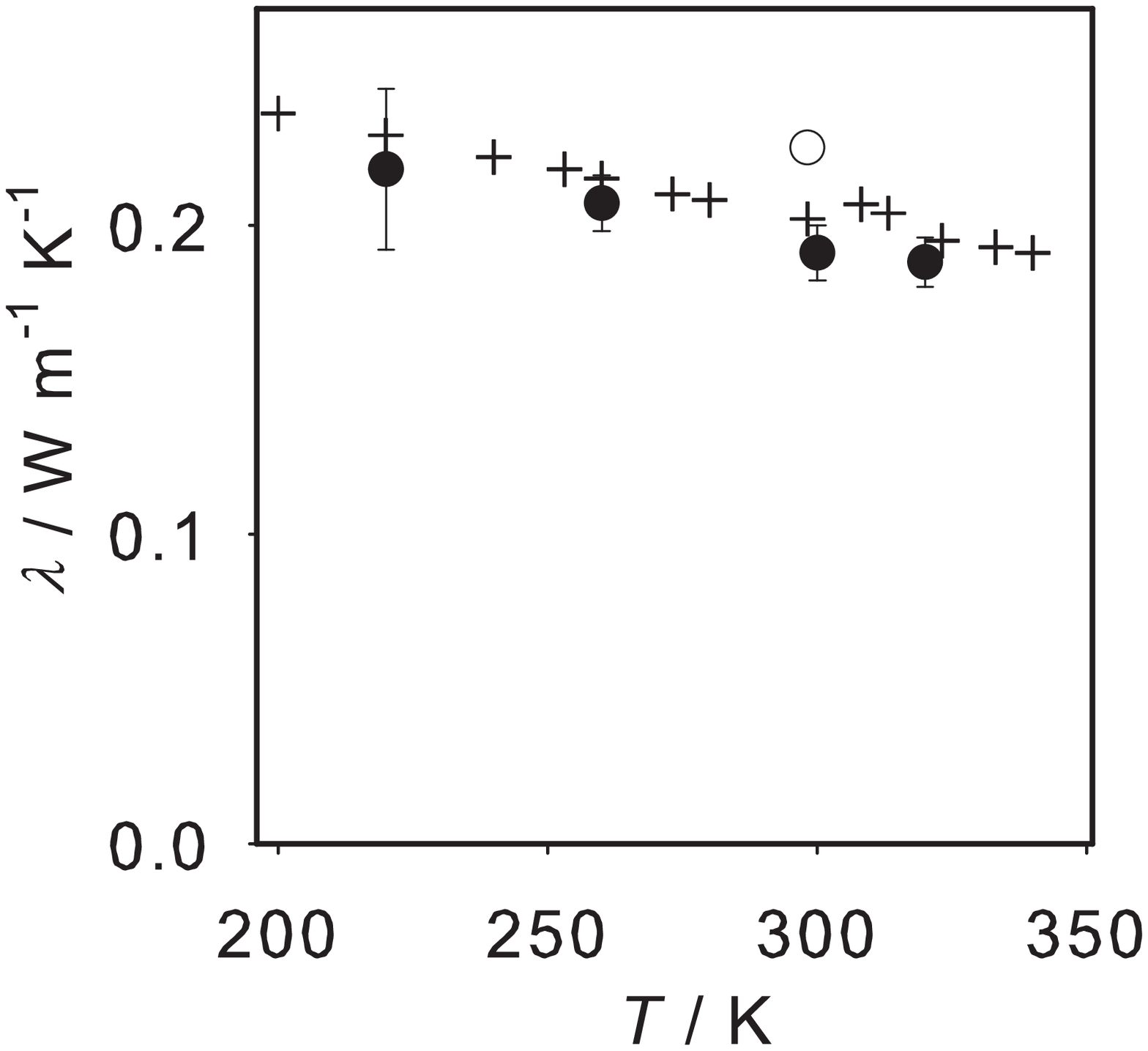}
\end{center}
\end{figure}

\begin{figure}[htb]
\begin{center}
\caption[Temperature dependence of the thermal conductivity of pure ethanol at $0.1$ MPa. Present simulation results ($\bullet$) are compared to experimental data \cite{vargaftik, poling, touloukian, CRC} ($+$) and to other simulation results \cite{petravic2005t} ($\Diamond$) and \cite{nieto2003t} ({\large$\circ$}).]{} \label{figtcetoh}
%\vskip1cm
\includegraphics[scale = 0.5]{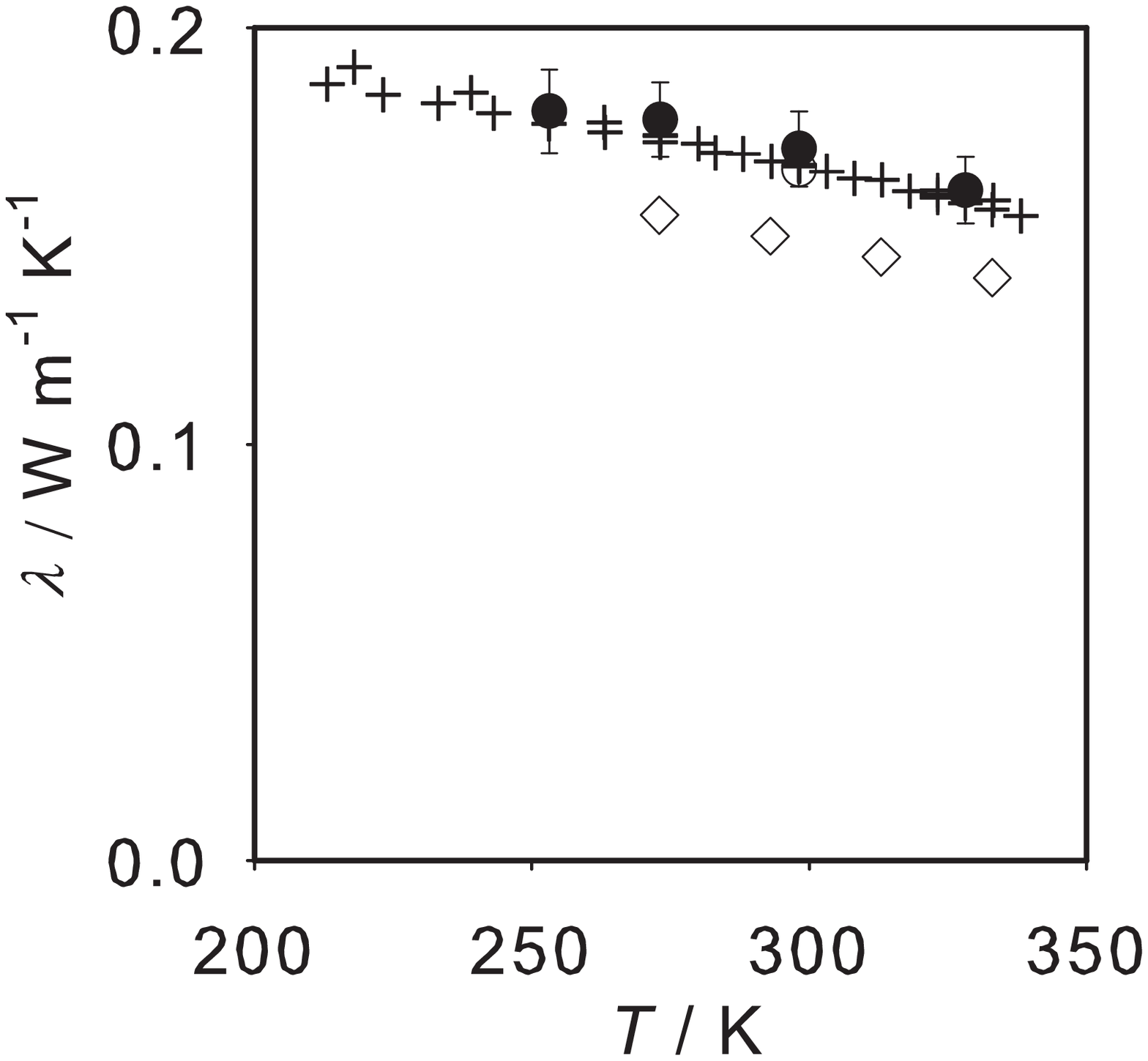}
\end{center}
\end{figure}

Present results for the thermal conductivity of pure methanol and pure ethanol at ambient pressure are summarized in Table \ref{resmeoh}. The agreement with the experimental data is remarkable with an average deviation of 5\% for methanol and of 2\% for ethanol, cf. Figures \ref{figtcmeoh} and \ref{figtcetoh}. The thermal conductivity for methanol is slightly underestimated by the present predictions, while for ethanol it is slightly overestimated. Previous results for the thermal conductivity of methanol \cite{nieto2003t} and ethanol \cite{petravic2005t}, using the OPLS models, show deviations of about $11$ and $10\%$, respectively. In contrast to present results, the calculated thermal conductivity by Petravi\'{c} \cite{petravic2005t} lies below the experimental data.

\subsection{Time dependent behavior of the Green-Kubo integrands}

\subsubsection*{Velocity autocorrelation function}

\begin{figure}[htb]
\begin{center}
\caption[Velocity autocorrelation function at 0.1 MPa of pure methanol at $180$ K ($-$) and $340$ K ($--$) as well as of pure ethanol at $173$ K ($-\cdot-$) and $333.15$ K ($\cdots$). ]{} \label{figvacf}
\vskip1cm
\includegraphics[scale = 0.7]{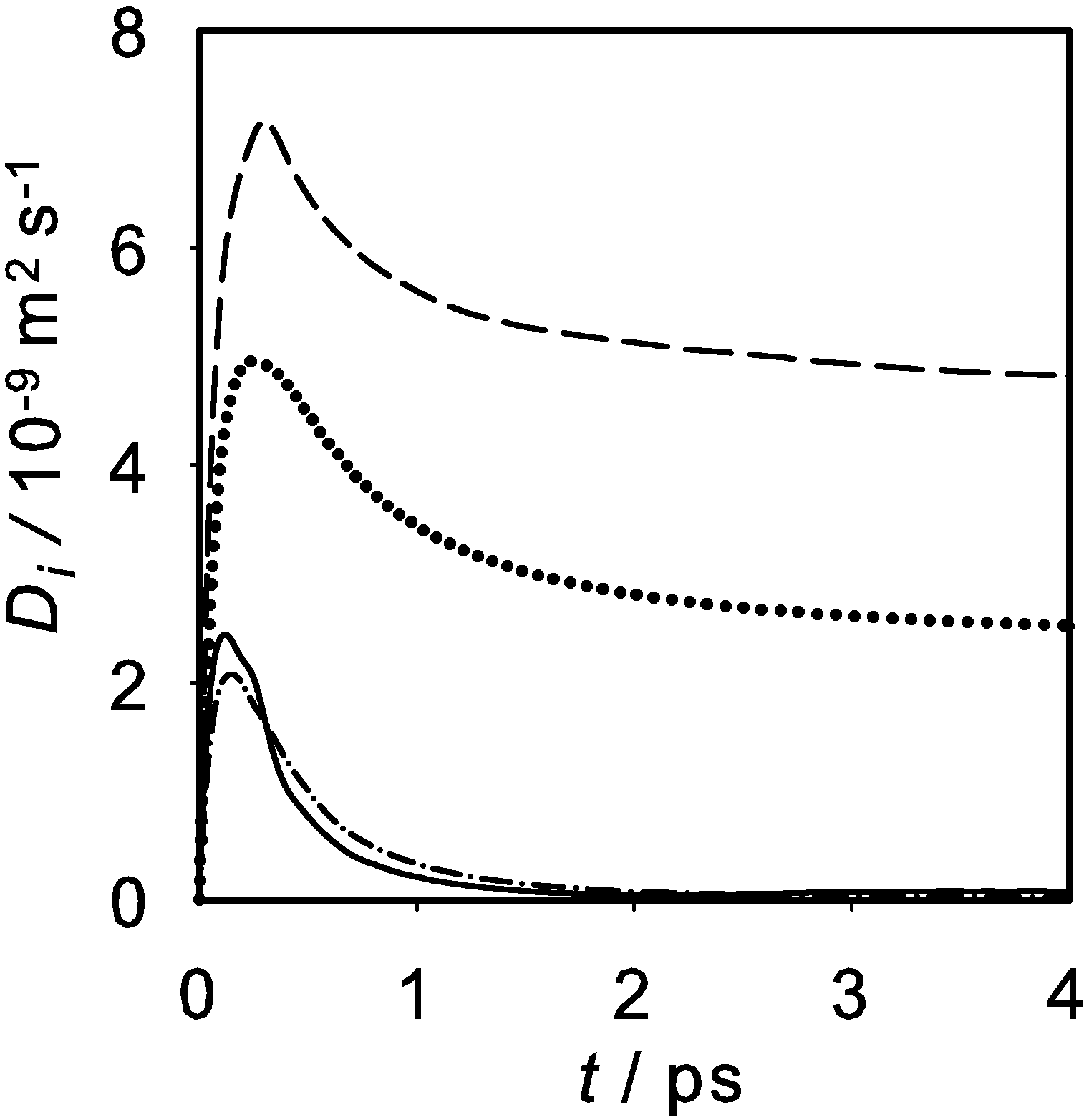}
\end{center}
\end{figure}

The velocity autocorrelation function provides an interesting insight into the liquid state. Selected normalized velocity autocorrelation functions (VACF) of pure methanol and ethanol are shown in Figure \ref{figvacf}. In agreement with findings in the literature \cite{alonso1991}, VACF may have two minima at short times ($t<0.5 ~\rm ps$). At low temperatures and high densities, the VACF of both pure liquids decrease sharply and become negative. After this first minimum, they increase slightly to drop again into a lower minimum, for methanol, or to a higher minimum, for ethanol. Beyond the second minimum, the VACF converge to zero following a characteristic path. The observed negative values of the VACF are related to backscattering collisions, also known as cage-effect, which govern short time dynamics. Here, the free movement of the molecules is hindered by surrounding molecules, which form a cage-like structure. Michels and Trappeniers \cite{michels1975} attributed this phenomenon to the formation of ``bound states'' since this effect can only be observed for molecular models that include attractive forces.

For higher temperatures at ambient pressure, where the liquid density is lower, molecular collisions are less frequent and the depth of the minima is gradually reduced until they almost disappear, cf. Figure \ref{figvacf}. At short times, the VACF decay faster at lower temperatures. Moreover, the initial decay of the VACF of methanol occurs earlier than in the corresponding VACF of ethanol. Methanol VACF exhibit a more pronounced backscattering oscillations than those of ethanol, particularly at low temperatures near its melting temperature ($T_{m}=175.37$ K) \cite{sindzigre1992}.

\begin{figure}[htb]
\begin{center}
\caption[Integral of the velocity autocorrelation function at 0.1 MPa of pure methanol at $180$ K ($-$) and $340$ K ($--$) as well as of ethanol at $173$ K ($-\cdot-$) and $333.15$ K ($\cdots$).]{} \label{figintvacf}
\vskip1cm
\includegraphics[scale = 0.5]{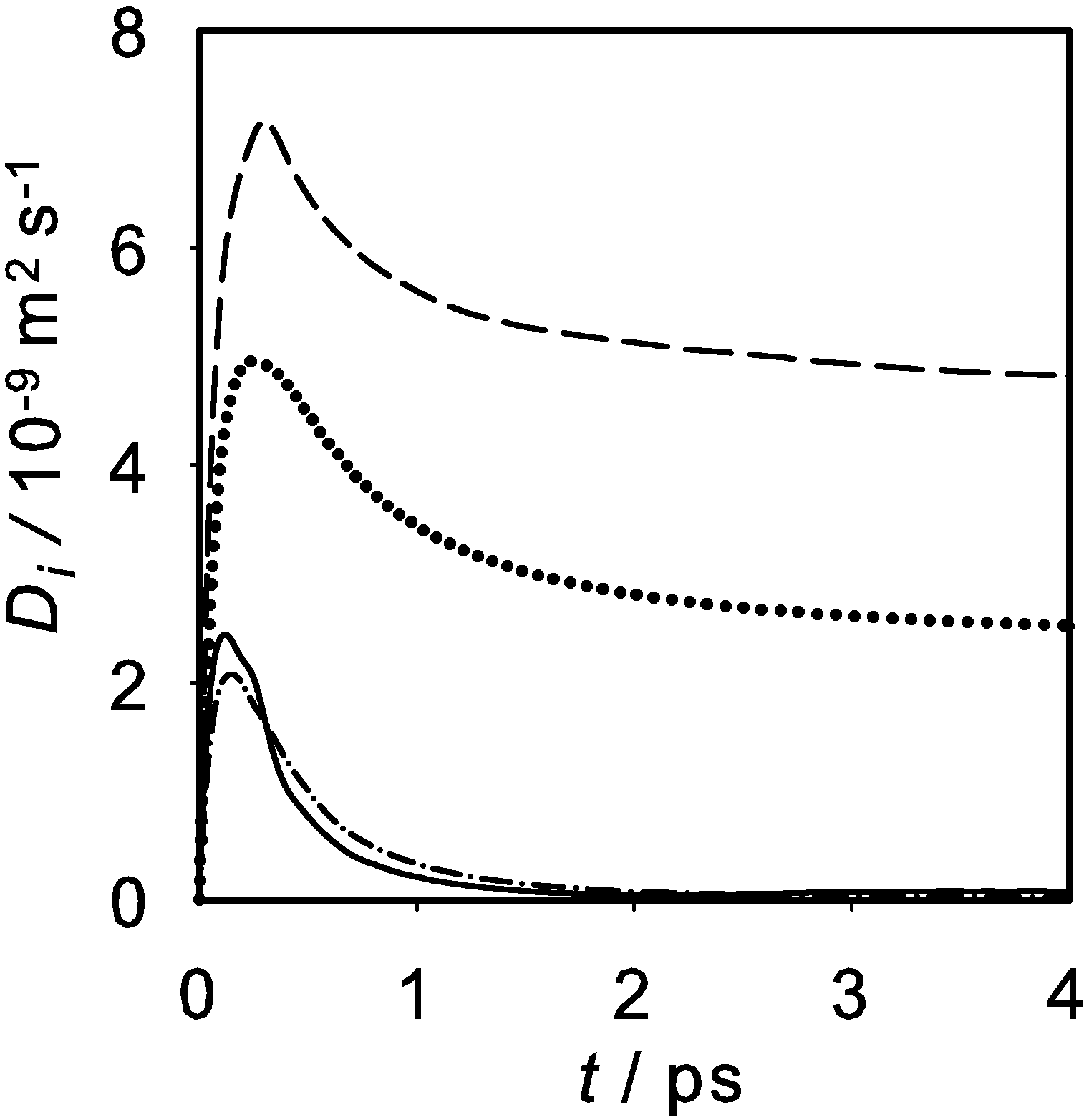}
\end{center}
\end{figure}

Figure \ref{figintvacf} shows the behavior of the integrals of the VACF given by Eq. (\ref{self}). It can be seen that the self-diffusion coefficient converges in all regarded cases to its final value after less than 4 ps. Thus, after this time interval the long-time tail of the VACF does not contribute significantly anymore.

\begin{figure}[htb]
\begin{center}
\caption[Velocity autocorrelation function of methanol in methanol (1) + ethanol (2) at 298.15 K and 0.1 MPa for different compositions: $x_{1} = 0.2~\rm mol~mol^{-1}$ ($\cdots$), $x_{1}=0.6~\rm mol~mol^{-1}$ ($-\cdot-$), $x_{1} = 0.8~\rm mol~mol^{-1}$ ($--$) and $x_{1} = 1~\rm mol~mol^{-1}$ ($-$).]{} \label{figvacfmixmeoh}
\vskip1cm
\includegraphics[scale = 0.5]{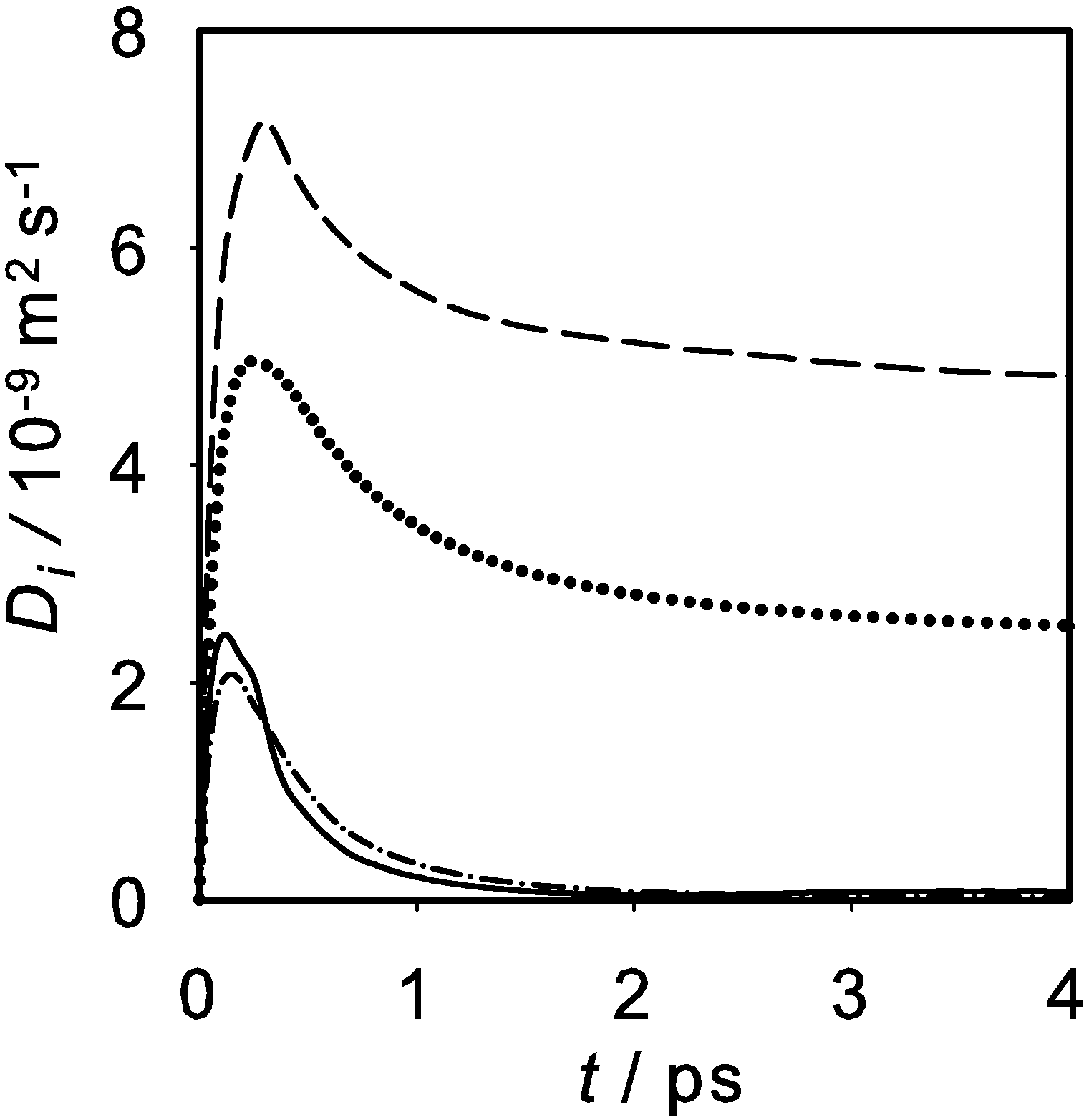}
\end{center}
\end{figure}

\begin{figure}[htb]
\begin{center}
\caption[Velocity autocorrelation function of ethanol in methanol (1) + ethanol (2) at 298.15 K and 0.1 MPa for different compositions: $x_{2} = 0.2~\rm mol~mol^{-1}$ ($\cdots$), $x_{2} = 0.4~\rm mol~mol^{-1}$ ($-\cdot-$), $x_{2} = 0.8~\rm mol~mol^{-1}$ ($--$) and $x_{2} =1~\rm mol~mol^{-1}$($-$).]{} \label{figvacfmixetoh}
%\vskip1cm
\includegraphics[scale = 0.5]{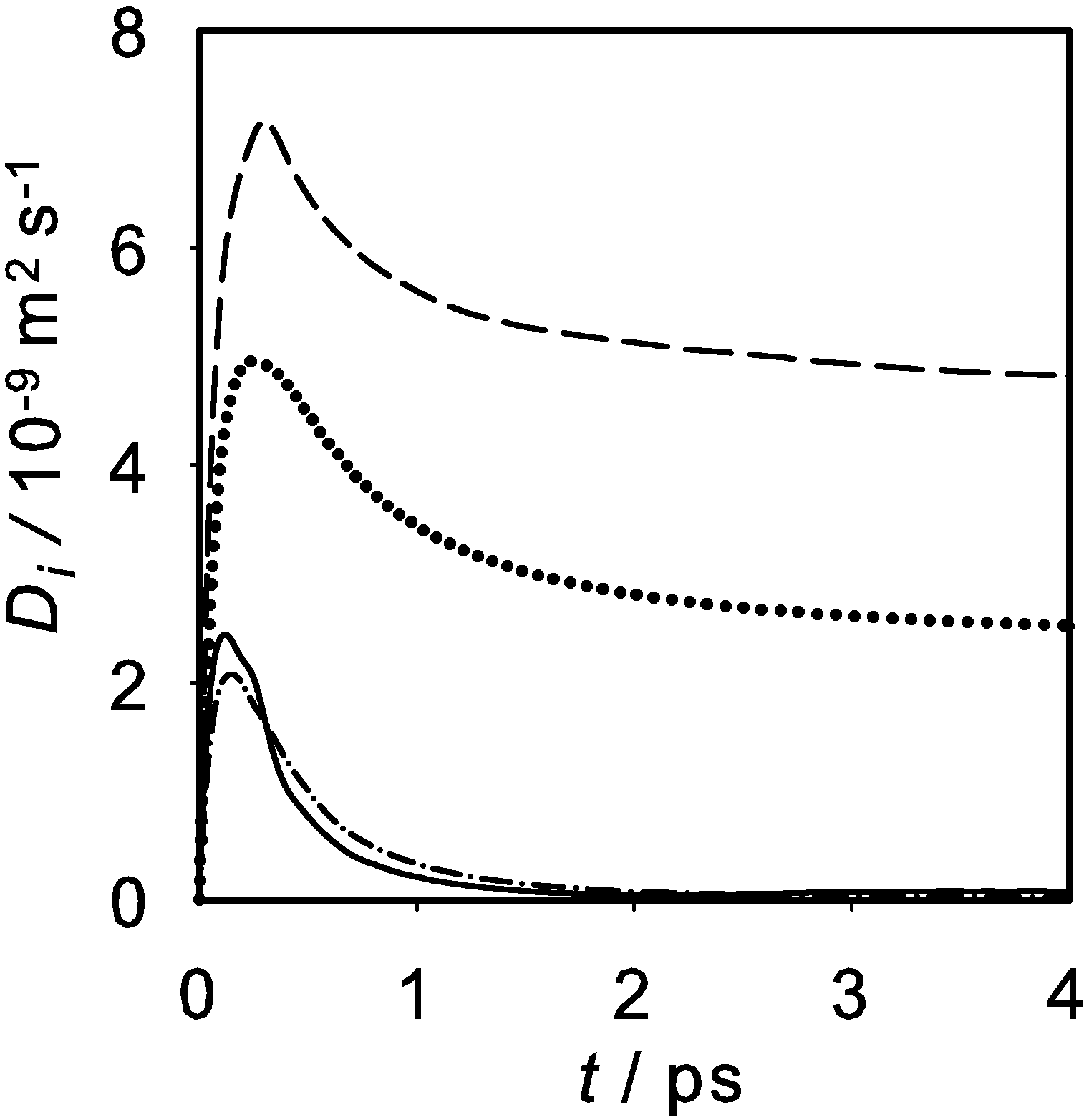}
\end{center}
\end{figure}

The behavior of the VACF of methanol and ethanol was also considered in the mixture. From Figures \ref{figvacfmixmeoh} and \ref{figvacfmixetoh} it can be seen that the VACF of methanol and ethanol in the mixture differ from those of the pure alcohols at the same temperature and pressure. Methanol VACF exhibit a significantly increased backscattering due to the presence of ethanol, cf. Figure \ref{figvacfmixmeoh}. This indicates that at short times, ethanol molecules further restrain the mobility of methanol molecules. Ethanol VACF also show a composition dependent behavior. In this case, the backscattering is reduced as the concentration of methanol increases, cf. Figure \ref{figvacfmixetoh}. This suggests that methanol molecules prevent the formation of cage-like structures here, allowing ethanol molecules to move more freely in the mixture.

\subsubsection*{Shear viscosity autocorrelation function}

Alder et al. \cite{alder1970} were the first to suggest the existence of a long-time tail in the shear viscosity autocorrelation function (SACF) near the phase boundary to solidification due to the low compressibility of such liquid states. Luo and Hoheisel \cite{luo1991, luo1991rev} showed that the long-time behavior of the SACF is determined primarily by long-lived orientational correlations, occurring in liquids containing anisotropic molecules. This behavior causes convergence problems in the Green-Kubo integral, since the SACF must be calculated over a relatively long time interval. 

\begin{figure}[htb]
\begin{center}
\caption[Shear viscosity autocorrelation function of pure methanol at 0.1 MPa and $260$ K ($-$), $278.15$ K ($-\cdot-$) and $340$ K ($--$).]{} \label{figsacf}
%\vskip1cm
\includegraphics[scale = 0.7]{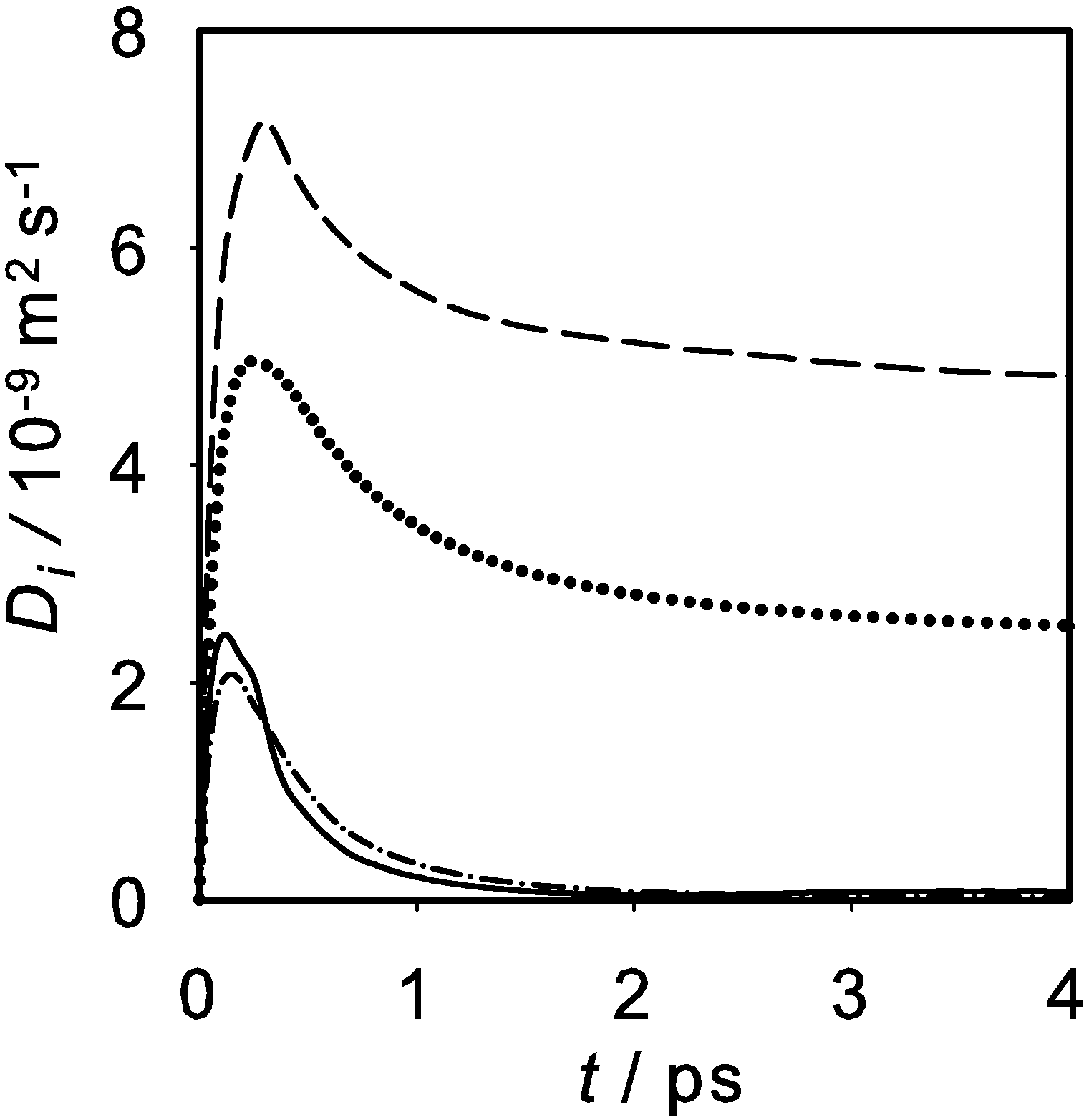}
\end{center}
\end{figure}

\begin{figure}[htb]
\begin{center}
\caption[Integral of the shear viscosity autocorrelation function of pure methanol at 0.1 MPa and $260$ K ($-$), $278.15$ K ($-\cdot-$) and $340$ K ($--$).]{} \label{figintsacf}
%\vskip1cm
\includegraphics[scale = 0.5]{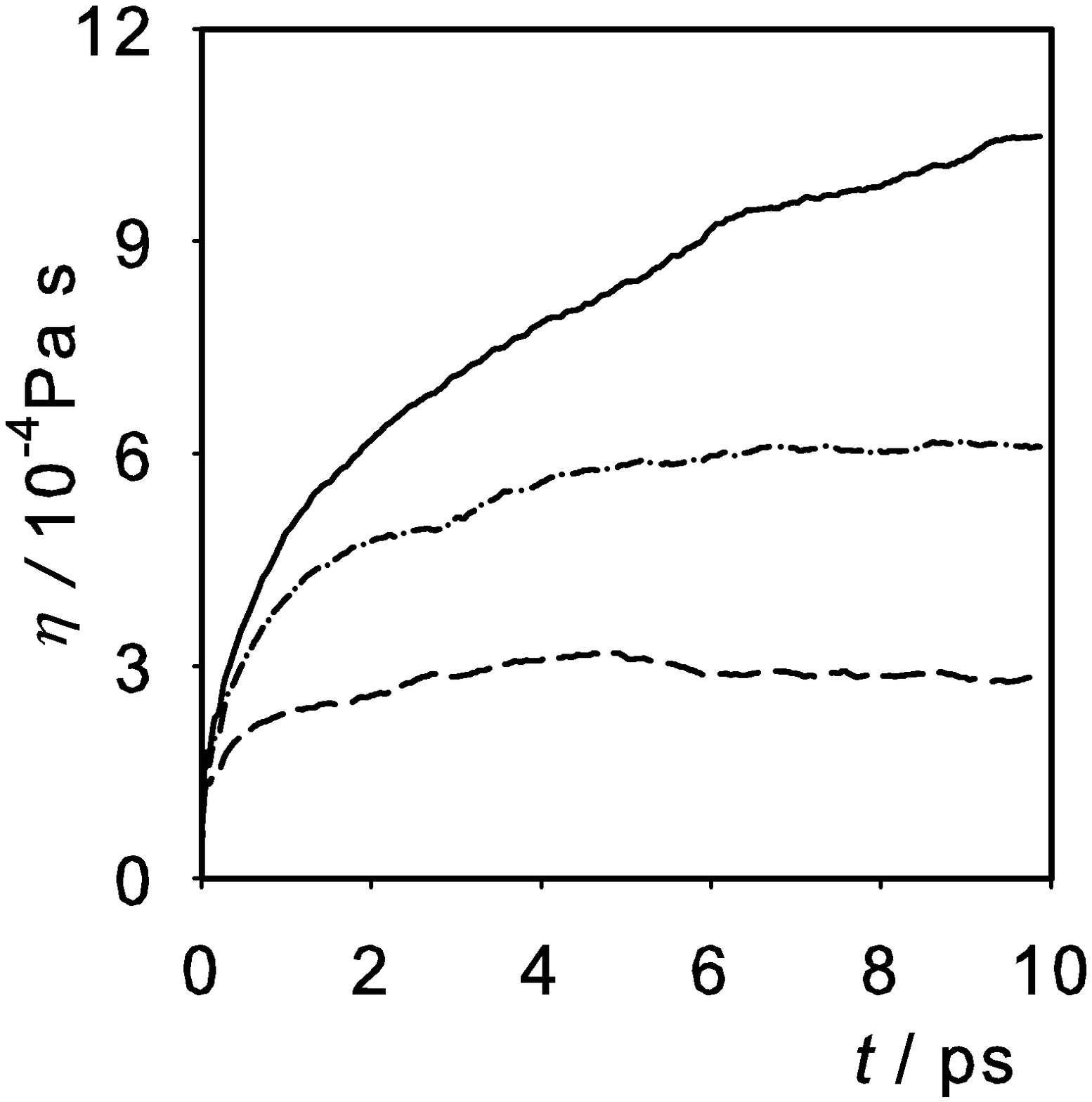}
\end{center}
\end{figure}

Figure \ref{figsacf} shows selected normalized SACF of pure methanol. The presence of a significant long-time tail can clearly be noticed in Figure \ref{figintsacf} where the integrals of the SACF of methanol are plotted for three temperatures. As expected, the behavior of the long-time tail of the SACF is highly dependent on molecule species and state point. For pure methanol at a low temperature of $260$ K, approximately $60\%$ of the contribution to the total shear viscosity comes from the first 2 ps. At a higher temperature of $340$ K the first 2 ps contribute $90\%$ to the total shear viscosity. In the case of ethanol (not shown here), the SACF are significant for even longer times, i.e. at $263$ K the first 2 ps of the SACF contribute only $45\%$ to the final value of the shear viscosity. At $338$ K this contribution reaches $75\%$. This difference can be explained by the higher molecular anisotropy of ethanol.

\begin{figure}[htb]
\begin{center}
\caption[Kinetic energy contribution of the shear viscosity autocorrelation function and its integral (inset) for pure methanol at $308$ K and 0.1 MPa.]{} \label{figsacfkk}
%\vskip1cm
\includegraphics[scale = 0.7]{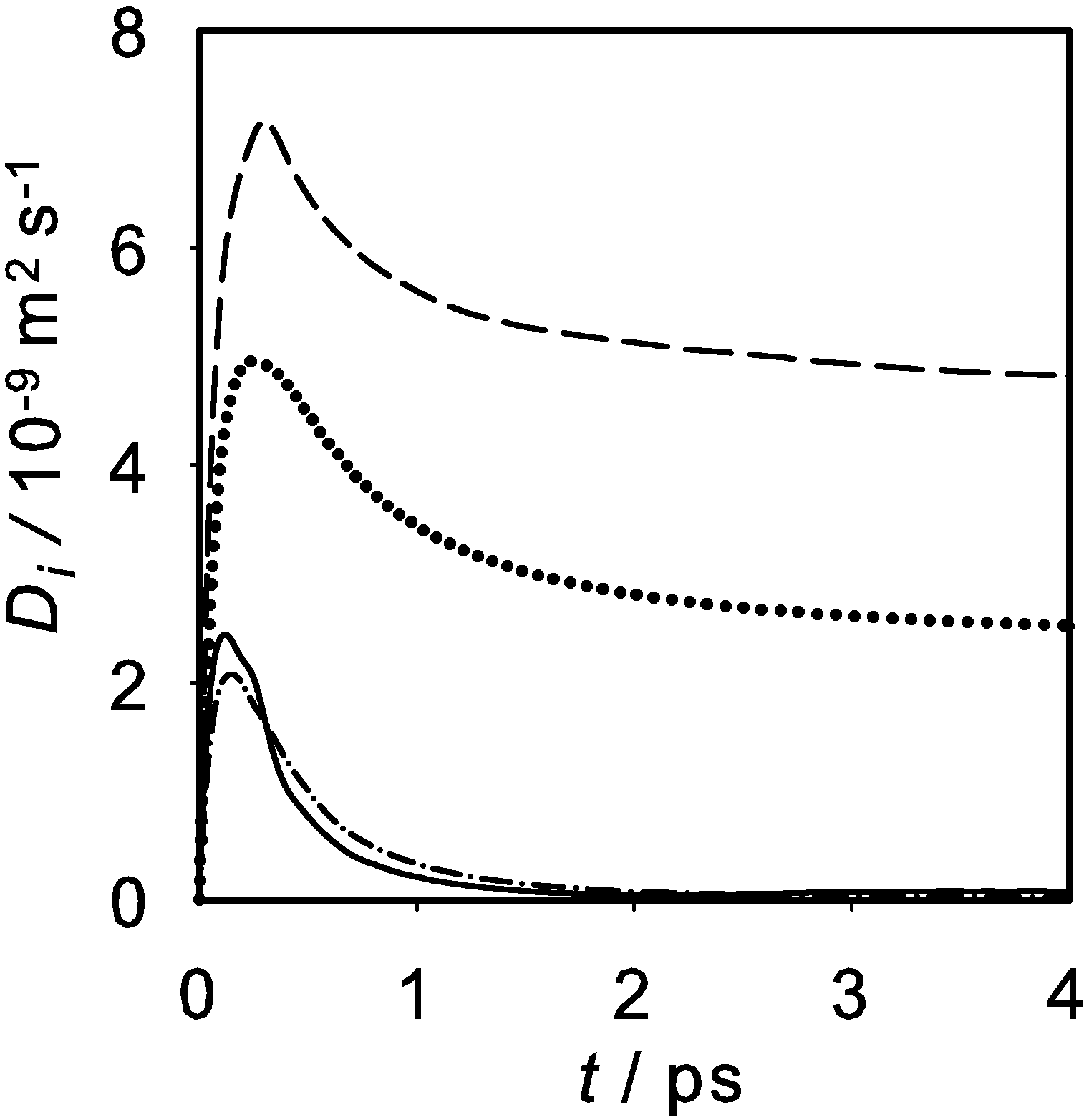}
\end{center}
\end{figure}

\begin{figure}[htb]
\begin{center}
\caption[Potential energy contribution to the shear viscosity autocorrelation function and its integral (inset) for pure methanol at $308$ K and 0.1 MPa.]{} \label{figsacfpp}
%\vskip1cm
\includegraphics[scale = 0.7]{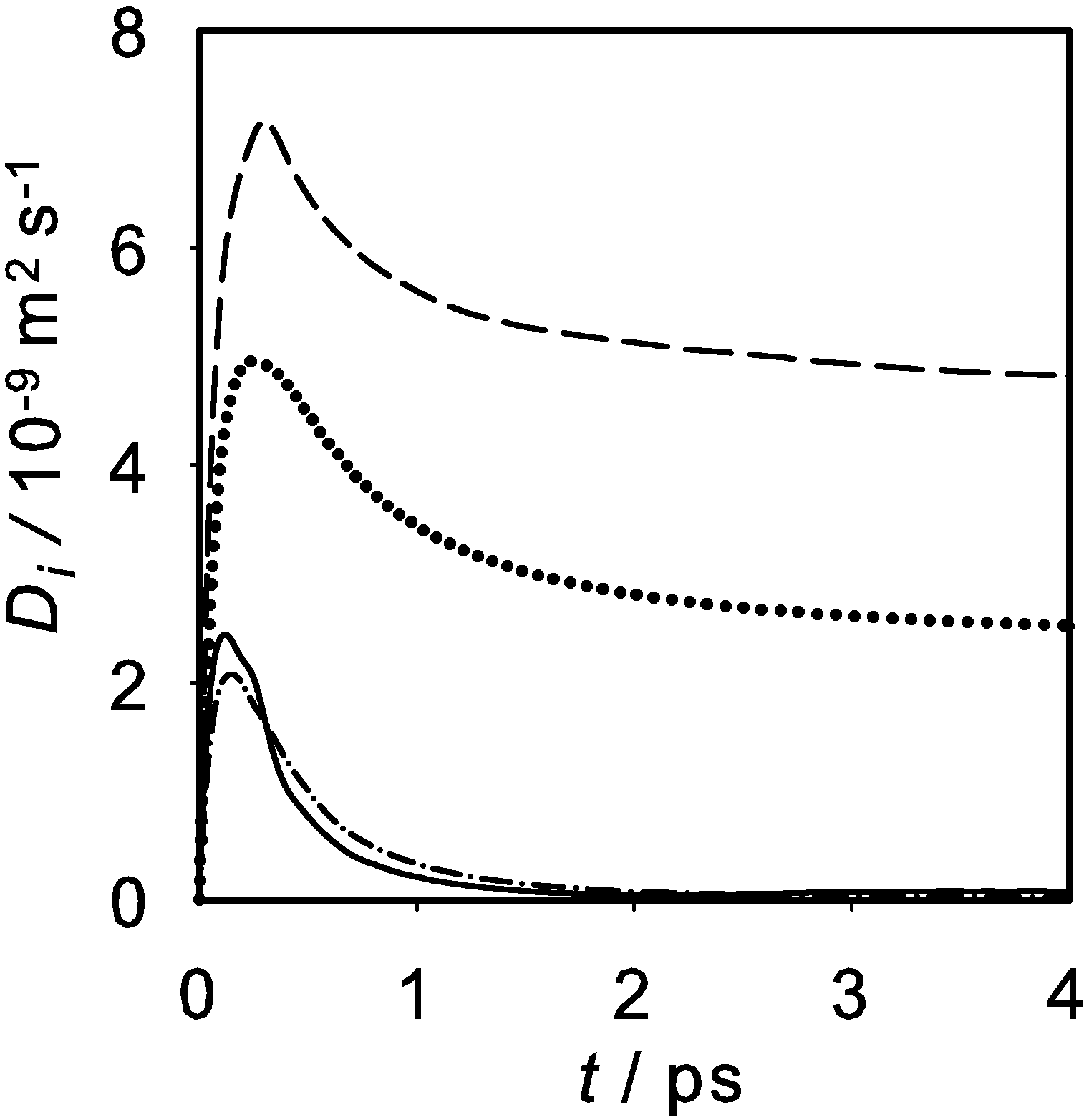}
\end{center}
\end{figure}

\begin{figure}[htb]
\begin{center}
\caption[Mixed kinetic-potential energy contribution to the shear viscosity autocorrelation function of pure methanol at 0.1 MPa and $260$ K ($-$), 278 K ($--$), 298 K ($-\cdot-$) and 308 K ($\cdots$).]{} \label{figsacfkp}
%\vskip1cm
\includegraphics[scale = 0.5]{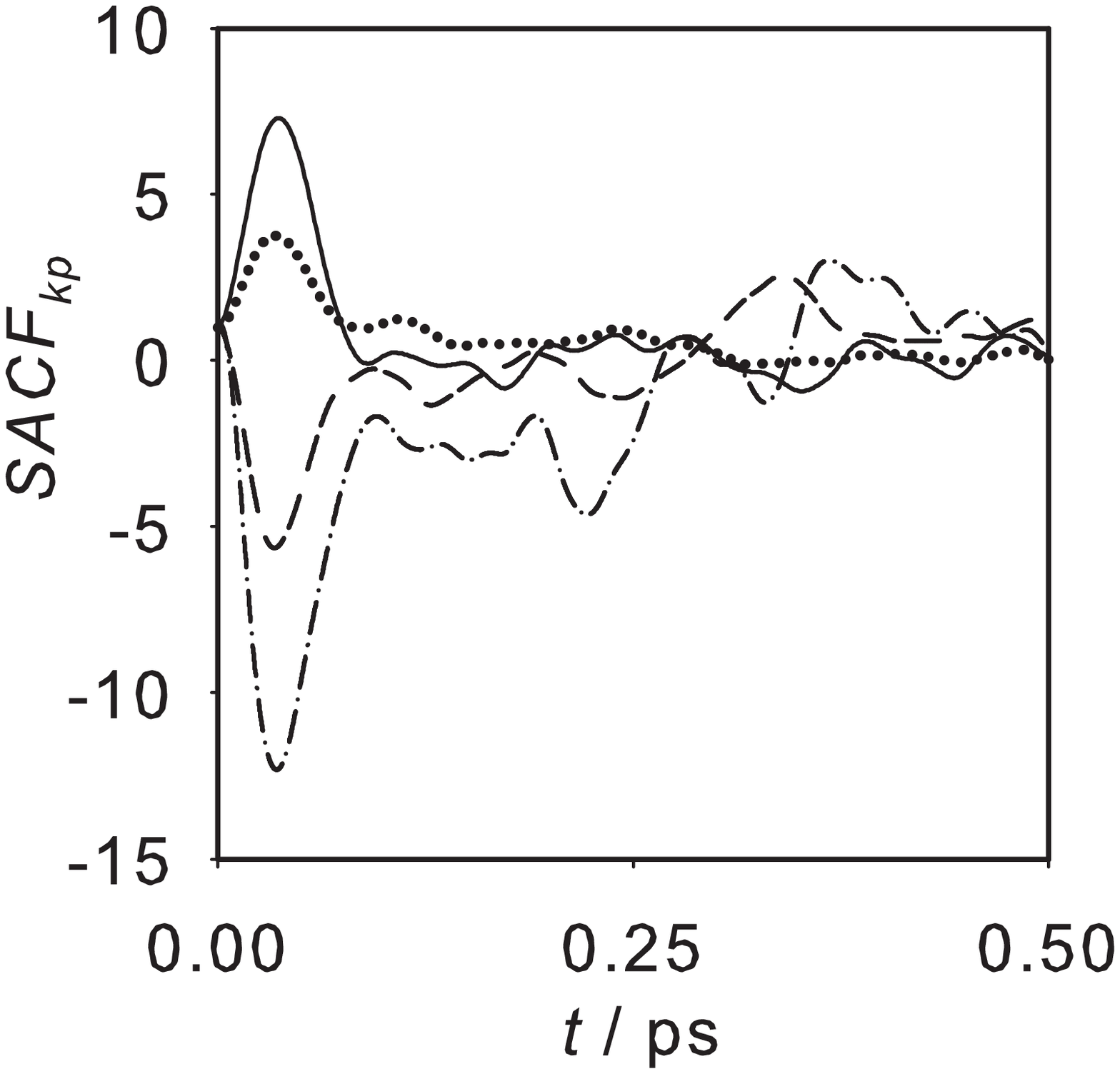}
\end{center}
\end{figure}

The kinetic, potential and mixed kinetic-potential energy contributions to the SACF were also considered in this work. Over the whole regarded range of state points, the behavior of the kinetic contribution is analogous to the monotonic chair-like decay of autocorrelation functions for non-attractive fluids. The corresponding time integral exhibits no important contribution from the long-time behavior, cf. Figure \ref{figsacfkk}. The potential energy contribution shows at short times oscillations similar to those of the VACF, cf. Figures \ref{figvacf} and \ref{figsacfpp}. These oscillations have been also related to the presence of "bound states" by Michels and Trappeniers \cite{michels1975} and by Meier et al. \cite{meier2004v}. The long-time tail is significant and approximates the behavior of the total SACF. The mixed kinetic-potential energy contribution shows a completely different behavior at short times, cf. Figure \ref{figsacfkp}, where data is presented for pure methanol at four temperatures. At very short times $(t< 0.1~\rm ps)$, two roughly mirrored patterns can be seen, the SACF decay/raise sharply to reach a minimum/maximum and oscillate strongly during their decay to zero. Meier et al. \cite{meier2004v}, who investigated simple LJ fluids, only observed initial positive peaks in the mixed kinetic-potential energy contribution. To our knowledge, the presence of negative initial peaks has not been reported before. The long-time tail behavior of the mixed contribution is difficult to analyze because of the presence of strong oscillations.

\subsection{Hydrogen bonding dynamics}

 Petravi\'{c} et al. \cite{petravic2003} have shown that internal degrees of freedom play a crucial role on the relaxation of the hydrogen bonding dynamics of liquid ethanol. For instance, they found that "freezing" bending and torsion in the OPLS model of ethanol produces a slowdown of the dynamics of the hydrogen bonds at $273$ K (they observed an increase of the hydrogen bond life time of about 27\%). For the methanol model used in this work, an excellent agreement between simulated and experimental site-site radial distribution functions and hydrogen bond statistics has been reported in previous work \cite{schnabel2007}. In the present work, the hydrogen bond dynamics of the rigid ethanol model was compared with the results using a flexible model version including bending and torsion. For this porpouse, a geometrical hydrogen bonding criterion with $r_{OH}$ below 2.6 \AA, $r_{OO}$ below 3.5 \AA~ and $\theta_{(OH...O)}$ below $30\,^{\circ}$ together with a life time probability accounting for intermittent hydrogen bonds \cite{nieto2003s} was used. In the flexible version, bending and torsion angles were allowed to evolve using a standard harmonic and a Fourier series potential, respectively. The potential constants were taken from the TraPPE intermolecular potential for alcohols \cite{chen2001}, keeping the equilibrium values of the bending and torsion angles of the rigid model and all the remaining inter-molecular model parameters as well as the bond distances between all sites. Simulations were then performed at a specified temperature of $273$ K (using the Nos\'e-Hoover thermostat) and a density of $17.62 ~\rm mol/\rm l$ to compute the radial distribution functions, the hydrogen bond dynamics and self-diffusion coefficient for both, rigid and flexible version of the present ethanol model.

\begin{figure}[htb]
\begin{center}
\caption[Comparison of the radial distribution functions obtained with the rigid ($-$) and flexible ($--$) model versions for ethanol at $273$ K and $17.62$ mol/l.]{} \label{figrdf}
%\vskip1cm
\includegraphics[scale = 0.7]{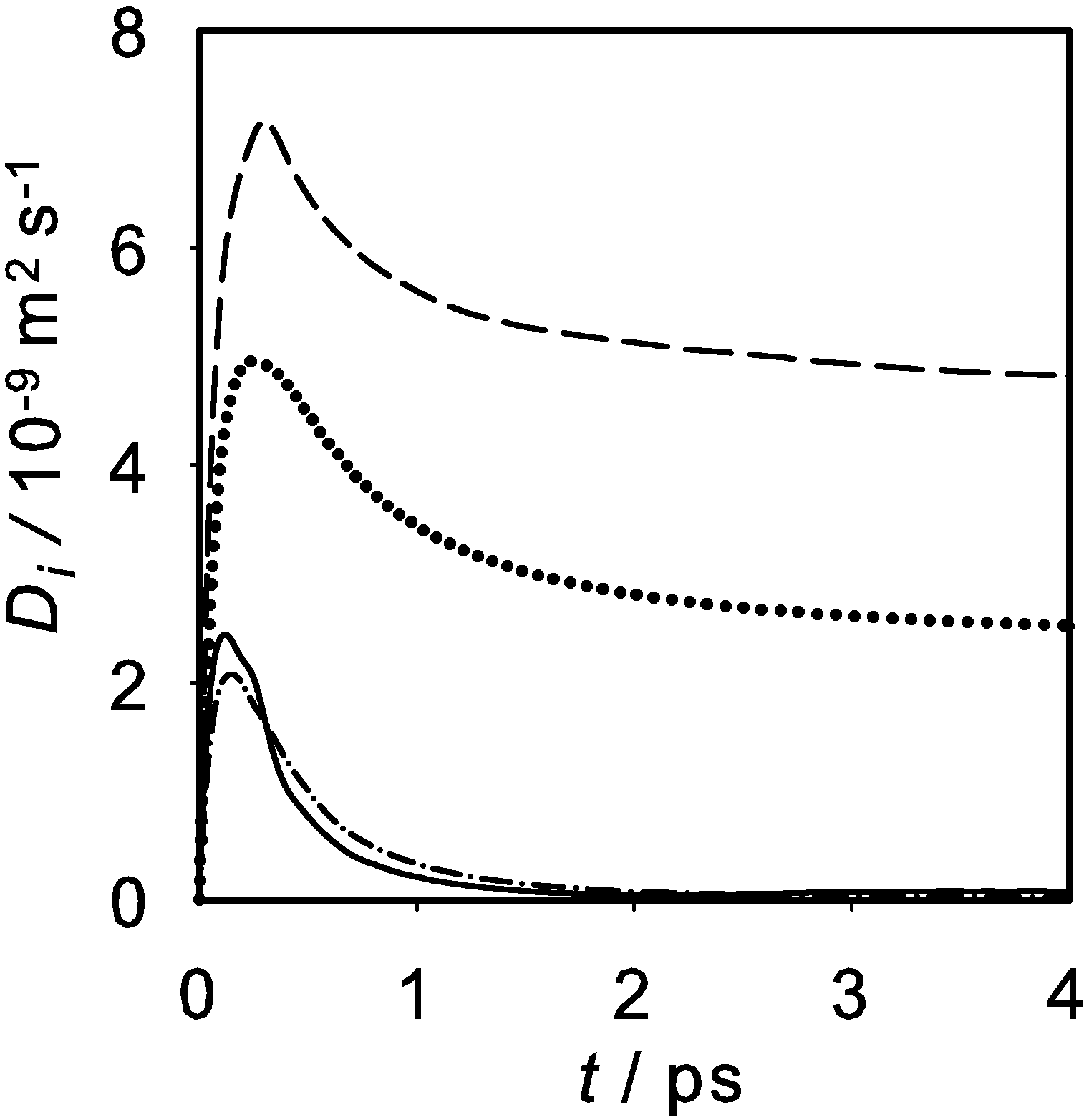}
\end{center}
\end{figure}

The comparison of the radial distribution functions from the flexible and the rigid model can be seen in Figure \ref{figrdf}. It was found that there are only minor differences in the liquid structure as revealed by the three radial distribution functions $g_{OO}$, $g_{OH}$ and $ g_{HH}$. The position and height of the first peaks are almost identical for both models, with the exception of $g_{HH}$, where a more pronounced peak is observed for the rigid model, cf. Figure \ref{figrdf}. Only small differences are noted around the second peak of $g_{OO}$ and $g_{OH}$, mainly related to the molecules hydrogen bonded in the second solvation shell. In the latter case, a more pronounced second peak was found for the rigid model. In general, it can be said that the inclusion of the flexibility does not strongly affect the structure of liquid ethanol around the first solvation shell.

In the next step, differences in the hydrogen bond dynamics for the two models were analyzed. Average life times of hydrogen bonds of $137$ and $80$ ps were obtained for the rigid and flexible versions, respectively. The flexibility significantly reduces the hydrogen bond life time by a factor of $1.7$. Petravi\'{c} et al. found in a similar investigation for the OPLS ethanol model a reduction in the hydrogen bond dynamics of about $1.3$. Hence, the effect is more pronounced for the present model. The hydrogen bond dynamics has a significant impact on the transport properties. Eg. at 273 K a self-diffusion coefficient of $0.802 \times 10^{-9} \rm m^{2}/\rm s$ was obtained for the flexible version, which is in perfect agreement with the values reported by Petravi\'{c} et al. \cite{petravic2003}. This value is higher by a factor of $1.6$ compared to the self-diffusion coefficient obtained for the rigid model ($0.493\times 10^{-9} \rm m^{2}/\rm s$) which is in excellent agreement with the experimental data ($0.48 \rm \times 10^{-9} \rm m^{2}/\rm s$) \cite{karger1990}. This comparison shows the impact of the internal degrees of freedom on the hydrogen bond dynamics and the self-diffusion coefficient. Similar effects can be expected also for other transport properties. Flexible and rigid models of H-Bonding systems obviously show significant differences regarding transport properties. This should also be kept in mind when assessing the comparisons of different models shown throughout this work.

\subsection{Vapor-liquid equilibria}

Vapor-liquid equilibria (VLE) often serve as a benchmark to assess molecular models. This is a useful choice, due to the fact that VLE represent the dominant feature in the region of fluid states. Indeed, they are nowadays preferably being used in the development of molecular models by optimizing the model parameters to VLE properties. This approach was also followed in the development of both regarded pure fluid models. In the next step, molecular models can be assessed on how they perform in mixtures. Based on experience for numerous binaries and ternaries, very good success can be expected, if the pure substance models are good.

Calculated VLE data of pure methanol and ethanol using present potential models have been reported previously for a temperature range from 270 to 490 K \cite{schnabel2007, schnabel2005}. The average deviations found for methanol in vapor pressure, saturated liquid density and heat of vaporization are of 1.1, 0.6, and 5.5\%, respectively \cite{schnabel2007}. The corresponding reported average deviations for ethanol are of 3.7, 0.3 and 0.9\%, respectively \cite{schnabel2005}.

\begin{figure}[htb]
\begin{center}
\caption[Vapor-liquid equilibria of the mixture methanol + ethanol at $393.15$ and $433.15$ K. Present simulation results ($\bullet$) are compared to experimental data \cite{butcher1966} ($+$). The lines represent the Peng-Robinson equation of state.]{} \label{figvle}
%\vskip1cm
\includegraphics[scale = 0.7]{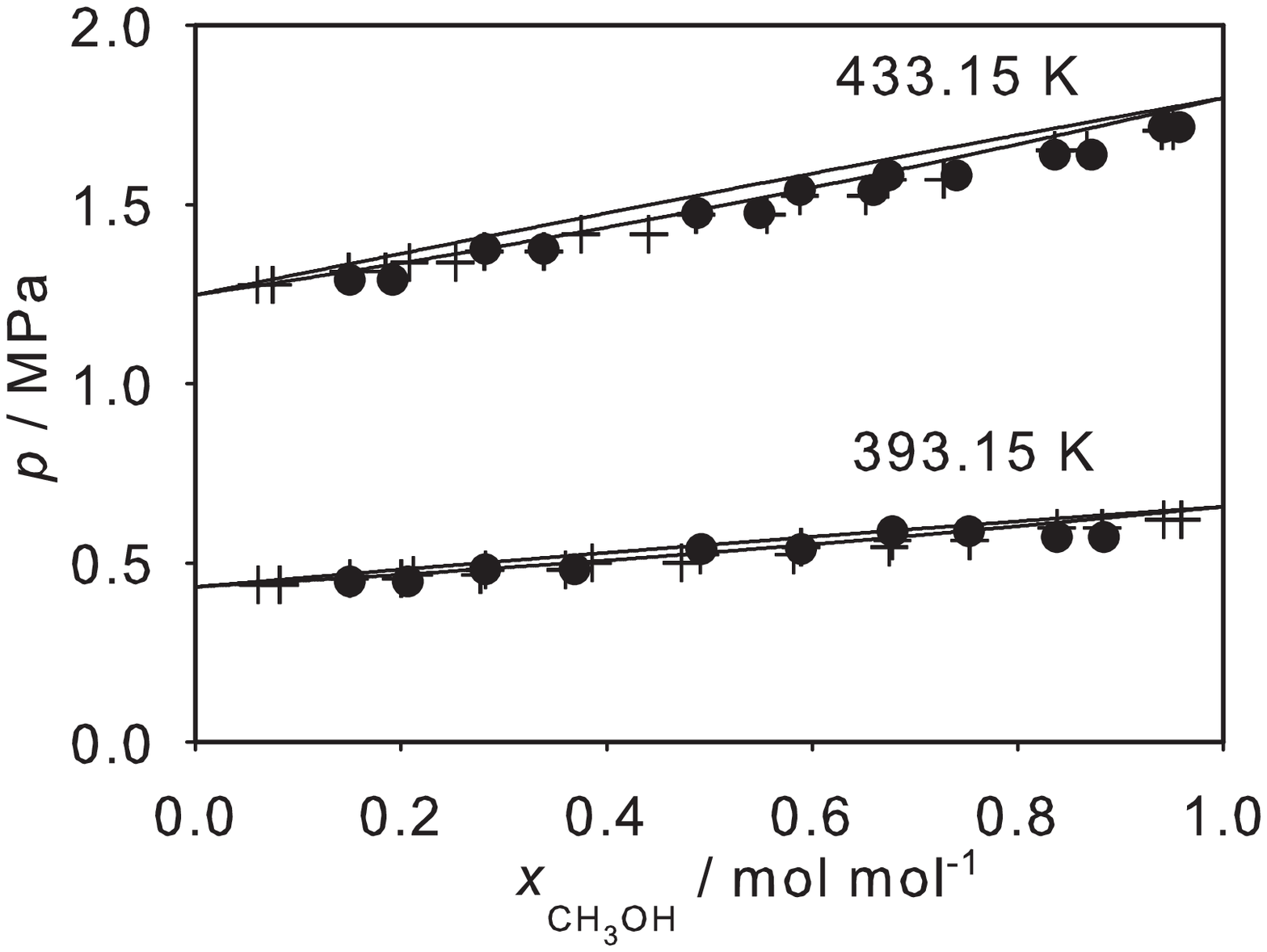}
\end{center}
\end{figure}

\begin{figure}[htb]
\begin{center}
\caption[Vapor to liquid mole fraction ratio of the mixture methanol + ethanol at $393.15$ and $433.15$ K. Present simulation results ($\bullet$) are compared to experimental data \cite{butcher1966} ($+$). The lines represent the Peng-Robinson equation of state.]{} \label{figvlexy}
%\vskip1cm
\includegraphics[scale = 0.7]{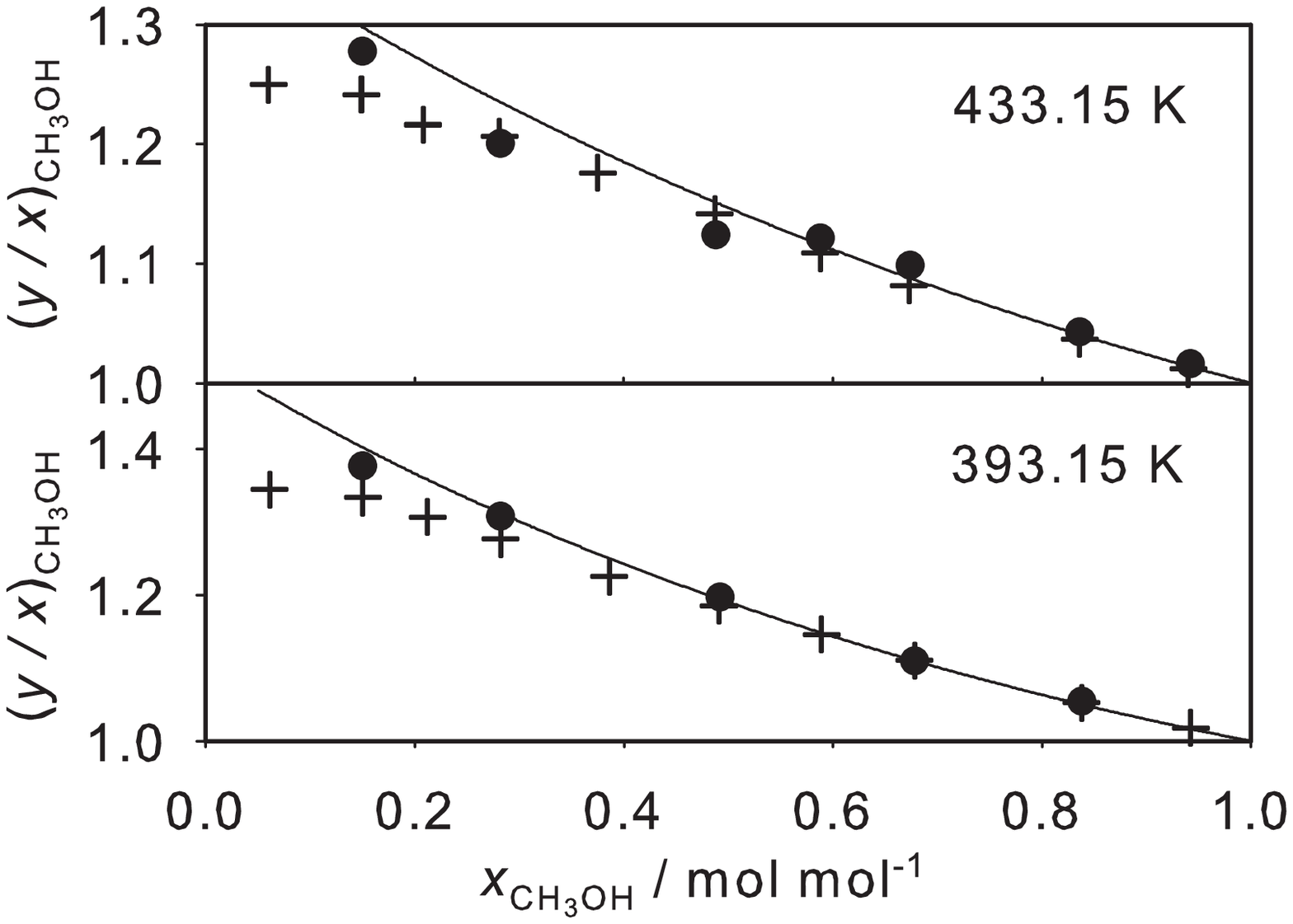}
\end{center}
\end{figure}

Present simulation results for the binary VLE are given in Table \ref{resvle} and compared in Figures \ref{figvle} and \ref{figvlexy} to experimental data by Butcher and Robinson \cite{butcher1966} and to the Peng-Robinson equation of state \cite{peng1976}. As in the molecular simulations no adjustment to binary data was carried out, also the Peng-Robinson equation was used without any adjustment to such data ($k_{ij}$~=~$0$). Figure \ref{figvle} shows the almost "ideal" behavior of methanol + ethanol for two temperatures. In fact, these were the highest two isotherms for which experimental data were available to us. At the higher one (443.15 K), however, it can be seen from the experimental data that the bubble line is not fully linear. The comparison of simulation and experiment is excellent throughout, both the pressure slope and the composition relation between vapor and liquid match good, only minor deviations are found in the ethanol-rich composition range where slightly too high vapor compositions are obtained.

\subsection{Excess properties}

Excess properties are basic mixing properties which contain real mixing behavior. In case of methanol + ethanol non-ideal mixing effects are very weak. At equimolar composition, where the mixing influence is strongest, experimental data suggest a volume increase of 0.017\% only. Also for enthalpy, a tiny positive excess contribution of 0.011 \% was found by experiment. This resolution is difficult to achieve by molecular simulation. 

To determine excess properties, a straightforward approach was used here. Three simulations at specified temperature and pressure were performed, two for the pure substances and one of the mixture at a given composition. A binary excess property given by

\bigskip
\begin{equation}\label{excess}
y^{\rm E}= y - x_{1}\cdot y_{1}-x_{2}\cdot y_{2} ,
\end{equation}

\noindent where $y^{\rm E}$ represents the excess volume $v^{\rm E}$ or the excess enthalpy $h^{\rm E}$. Enthalpies or volumes of the pure liquids methanol and ethanol as well as of their binary mixture are denoted by $y_{1}$, $y_{2}$ and $y$, respectively. 

\begin{figure}[htb]
\begin{center}
\caption[Composition dependence of the excess volume for the mixture methanol + ethanol at $298.15$ K and 0.1 MPa. Present simulation results ($\bullet$) are compared to experimental data \cite{benson1970, ogawa1987, zarei2007} ($+$).]{} \label{figve}
%\vskip1cm
\includegraphics[scale = 0.5]{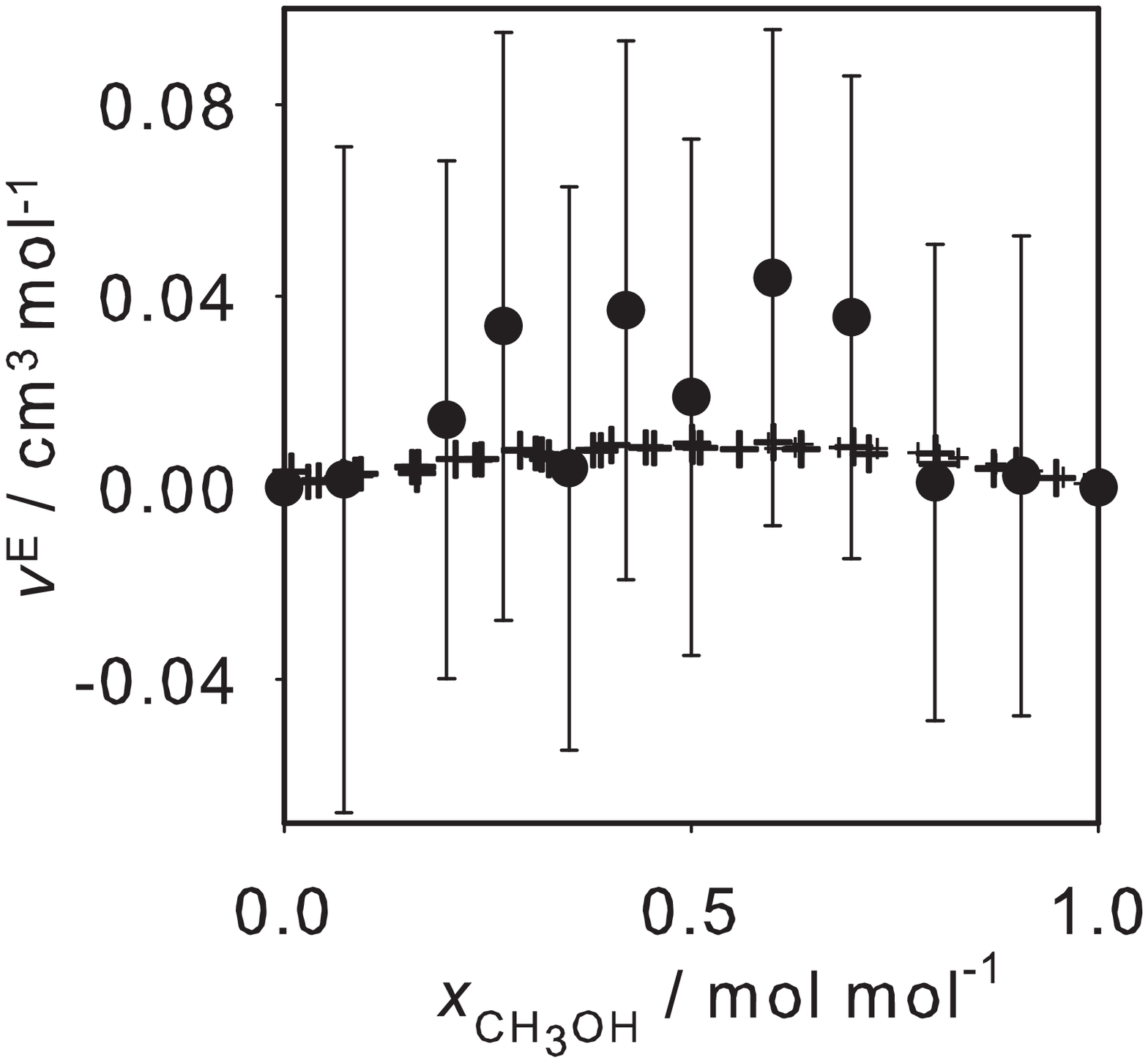}
\end{center}
\end{figure}

\begin{figure}[htb]
\begin{center}
\caption[Composition dependence of the excess enthalpy for the mixture methanol + ethanol at $298.15$ K and 0.1 MPa. Present simulation results ($\bullet$) are compared to experimental data \cite{pflug1968} ($+$).]{} \label{fighe}
%\vskip1cm
\includegraphics[scale = 0.5]{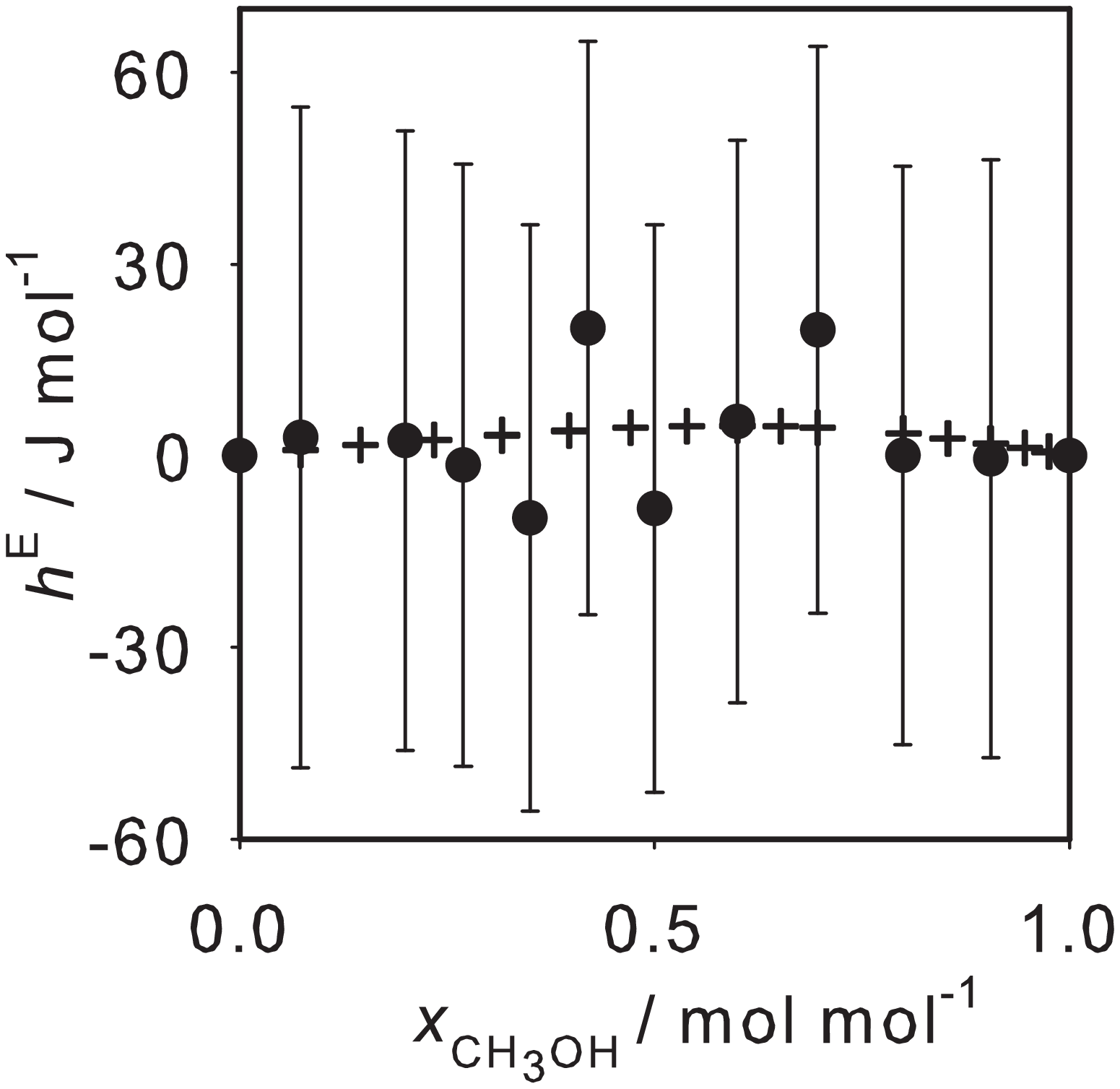}
\end{center}
\end{figure}

Table \ref{resexcess} contains present simulation data for ambient conditions which are compared to experimental results \cite{benson1970, ogawa1987, zarei2007, pflug1968} in Figures \ref{figve} and \ref{fighe}. The excess volume, cf. Figure \ref{figve}, agrees with the experiment in all cases within its seemingly large error bars. However, it should be noted that the simulation uncertainty is only around 0.1\% of the total volume of the mixture. A slight tendency of the simulation data towards higher excess volumes might be assumed. In Figure \ref{fighe}, the present excess enthalpy is shown together with experimental data. The maximal excess contribution from experiment is only 0.011\%, which is well below the simulation uncertainty of around 0.1\%. By inspection of the simulation data set as a whole, an even better agreement with experiment can be assumed.

\section{Conclusion}
Transport and mixture properties for methanol and ethanol were predicted by molecular dynamics simulation on the basis of rigid, non-polarizable molecular models from preceding work. Self- and Maxwell-Stefan diffusion coefficients as well as shear viscosity were calculated by EMD and the Green-Kubo formalism. Comparing to experimental data for the two pure alcohols, present self-diffusion coefficient and shear viscosity data are very good to excellent over a temperature range of about 100 K. Slightly higher deviations were found for the shear viscosity at low temperatures, which are mainly due to the long-time behavior of the shear viscosity autocorrelation function. Regarding the mixture methanol + ethanol at ambient conditions over the full composition range, both self-diffusion coefficients also excellently agree with experimental data. The same holds for shear viscosity, especially in the methanol-rich region. Maxwell-Stefan diffusion coefficient data is, to our knowledge, not available in the literature. Present predictions correspond very well with the model of Darken. The thermal conductivity was determined by BD reverse NEMD. Over a temperature range of about 100 K, an excellent agreement with the experiment was found as well for both pure fluids. Furthermore, predicted VLE data for the mixture methanol + ethanol along two isotherms correspond with experimental data practically throughout within the simulation uncertainty. The same holds for excess volume and excess enthalpy at ambient conditions in the full composition range. It has been demonstrated that present rigid, non-polarizable models for ethanol and methanol show a better performance to reproduce transport properties when compared to standard rigid and flexible models available in the literature.

\clearpage

%\subsection*{Simulation Details}
\section*{Appendix~~~Simulation details \label{simdet}}

In this work, EMD simulations for transport properties were made in two steps. In the first step, a short simulation in the isobaric-isothermal ($NpT$) ensemble was performed at ambient pressure to calculate the density at the desired temperature. In the second step, a canonic ($NVT$) ensemble simulation was performed at this temperature and density, to determine the transport properties. The simulations were carried out in a cubic box with periodic boundary conditions containing $2048$ molecules. Newton's equations of motion were solved using a fifth-order Gear predictor-corrector numerical integrator. The temperature was controlled by velocity scaling. In all EMD simulations, the integration time step was $0.98$ fs. The cut-off radius was set to $r_{c}= 15$ \AA. Electrostatic long-range corrections were made using the reaction field technique with conducting boundary conditions $(\epsilon_{RF}=\infty)$. On the basis of a center of mass cut-off scheme, the LJ long-range interactions were corrected using angle averaging \cite{lustig1988}. The simulations were equilibrated in the $NVT$ ensemble over $10^{5}$ time steps, followed by production runs of $0.5$ to $2 \times10^{6}$ time steps. Diffusion coefficients and shear viscosity were calculated using Eqs. (\ref{self}) to (\ref{shear}) with up to $9~900$ independent time origins of the autocorrelation functions. The sampling length of the autocorrelation functions varied between $9$ and $14$ ps. The separation between time origins was chosen so that all autocorrelation functions have decayed at least to ${1}/{e}$ of their normalized value to guarantee their time independence \cite{schoen1984}. The uncertainties of the predicted values were estimated using a block averaging method \cite{allen1987}.

In the NEMD simulations with the PeX algorithm for the thermal conductivity, 800 molecules were placed in a parallelepiped box with $L_{z} = 3~ L_{x} = 3~ L_{y}$, where $L_{i}$ is the length of the simulation box in the different spatial directions. Periodic boundary conditions were applied in all directions. The system was then equilibrated over 1 ns at the desired temperature and pressure by $NpT$ simulation using a weak coupling bath \cite{berendsen1984} with long-range corrections \cite{allen1987} for pressure and energy. The resulting density of the equilibration run was then employed to generate a new set of simulations in order to develop the thermal gradient in a run of 2 ns using the NEMD scheme. In this case an exchange frequency $\upsilon = 3.3~\rm fs^{-1}$ (i.e. every $150$ time steps) for the energy flux was used. Electrostatic interactions were treated with the reaction field with conducting boundary conditions. The thermal conductivity was then computed as the average of 4 different runs over 1 ns. The integration of the equations of motion was performed with a time step of 2 fs with a cut-off radius of 12 \AA. A Verlet neighbor list was employed to improve the performance of the simulations. 

The VLE calculations of the mixture methanol + ethanol were performed using the Grand Equilibrium method \cite{vrabec2002g} in combination with the gradual insertion technique \cite{nezbeda1991}. The latter, a Monte Carlo based expanded ensemble method, was applied to determine the chemical potential in the liquid phase. In comparison to Widom's test particle method \cite{widom1963}, where a full molecule is inserted into the fluid, the gradual insertion method introduces a fluctuating molecule. This molecule undergoes changes in a set of discrete states of coupling with all other real molecules of the fluid. Preferential sampling was performed in the vicinity of the fluctuating molecule to improve accuracy. The gradual insertion simulations were done with 500 molecules in the liquid phase. Starting from a face-centered cubic lattice arrangement, the simulation runs were equilibrated over $2~000$ cycles in the $NpT$ ensemble. The production phase was performed over $30~000$ Monte Carlo cycles. Further simulation parameters of these runs were taken from Vrabec et al. \cite{vrabec2002}.

To evaluate the excess properties, EMD simulations in the $NpT$ ensemble were performed. The system contained 1372 molecules in a cubic box with periodic boundary conditions. The molecular trajectories were sampled with the algorithms and parameters as described above. The simulations were equilibrated in the $NpT$ ensemble over $80~000$ time steps, followed by a production run over $600~000$ time steps.

\clearpage

%\bibliographystyle{jpc}
%\bibliography{bibliosim2,biblioexp}

\clearpage

%Literature
\begin{table}[ht]
\begin{center}
\caption[]{Literature review regarding transport properties of methanol, ethanol and their binary mixture calculated by molecular simulation. \label{literature}}
\bigskip
\begin{tabular}{l| c c c}
\hline \hline
&&&\\[-2.0ex]
   & Self-diffusion  & Shear    & Thermal  \\
&           coefficient     & viscosity& conductivity\\
\hline
Methanol  & \cite{lucassen1999, haughney1987,pereira2001, casulleras1992, asahi1998} & \cite{wheeler1997} &  \cite{nieto2003t}\\
Ethanol  & \cite{petravic2003, shihao2007, taylor2003, saiz1997, sandeep2005, gonzalez1999, pereira2001} & \cite{petravic2005v, zhao2007} & \cite{petravic2005t, nieto2003t}\\
Methanol + Ethanol & $-$ & $-$ & $-$ \\
\hline \hline
\end{tabular}
\end{center}
\end{table}
\clearpage

% Models
\begin{table}[htb]
\caption{Lennard-Jones, point charge and geometric parameters of the molecular models for methanol and ethanol, cf. Eq. (\ref{equ_pot}). Boltzmanns's constant is $k_{\rm B}$ and the electronic charge is $e$.\label{tab_meoh_pot}}
\bigskip
\begin{tabular}{|l|c|c|c|} \hline
Site                  & $\sigma_{aa}$ & $\epsilon_{aa}/k_{\rm B}$ & $q_{ia}$\\
                      & \r{A}         & K                         & $ e $\\ \hline
\multicolumn{4}{|l|}{Methanol}\\ \hline
$\rm{S}_{\rm{CH3}}$   & 3.7543        & 120.592                   & \multicolumn{1}{|r|}{$+$0.24746}\\
$\rm{S}_{\rm{OH}} $   & 3.0300        &\087.879                   & \multicolumn{1}{|r|}{$-$0.67874}\\
$\rm{S}_{\rm{H}}  $   &$-$            &$-$                        & \multicolumn{1}{|r|}{$+$0.43128}\\ \hline
\multicolumn{4}{|l|}{Ethanol}\\ \hline
$\rm{S}_{\rm{CH3}}$   & 3.6072      & 120.15                &    $-$       \\
$\rm{S}_{\rm{CH2}}$   & 3.4612      & 86.291                & \multicolumn{1}{|r|}{$+$0.25560} \\
$\rm{S}_{\rm{OH}} $   & 3.1496      & 85.053                & \multicolumn{1}{|r|}{$-$0.69711}  \\
$\rm{S}_{\rm{H}}  $   & $-$         & $-$                   & \multicolumn{1}{|r|}{$+$0.44151}   \\ \hline
\end{tabular}
\end{table}
\clearpage

% Results Methanol and Ethanol
\begin{table}[ht]
\begin{center}
\caption[]{Density, self-diffusion coefficient, shear viscosity and thermal conductivity of pure liquid methanol and ethanol at $0.1$ MPa from present molecular dynamics simulation. The number in parenthesis indicates the statistical uncertainty in the last digit.\label{resmeoh}}
\bigskip
\begin{tabular}{l @{\hspace{0.5cm}} c @{\hspace{0.7cm}} c @{\hspace{0.7cm}} c@{\hspace{0.7cm}} c}
\hline \hline
&&&&\\[-2.0ex]
 $T$         &$\rho$                  &$ D_{i}$                   & $\eta$                    &$\lambda$\\
 $ \rm K $   &$\rm mol\0\rm l^{-1}$   &$\rm{10^{-9}m^{-2}s^{-1}}$ & $10^{-3} \rm Pa\0\rm s$   &$\rm W m^{-1} K^{-1}$\\
\hline
\multicolumn{5}{l}{Methanol}\\
\hline
213      & 27.02   & \multicolumn{1}{l}{0.318 (5)}     & \multicolumn{1}{l}{3.1\0 (4)}  &           \\ 
220      & 26.80   & \multicolumn{1}{l}{0.433 (6)}     & \multicolumn{1}{l}{2.0\0 (2)}  & 0.22\0 (3) \\
240      & 26.22   & \multicolumn{1}{l}{0.750 (8)}     & \multicolumn{1}{l}{1.7\0 (2)}  &           \\ 
260      & 25.63   & \multicolumn{1}{l}{1.21\0 (1)}    & \multicolumn{1}{l}{1.0\0 (2)}  & 0.207 (9) \\ 
278.15   & 25.14   & \multicolumn{1}{l}{1.72\0 (1)}    & \multicolumn{1}{l}{0.64 (8)}   &           \\ 
288      & 24.81   & \multicolumn{1}{l}{2.09\0 (2)}    & \multicolumn{1}{l}{0.62 (7)}   &           \\ 
298.15   & 24.51   & \multicolumn{1}{l}{2.41\0 (2)}    & \multicolumn{1}{l}{0.57 (6)}   &           \\
300      & 24.50   &                                   &                                & 0.191 (9) \\
308.15   & 24.19   & \multicolumn{1}{l}{2.97\0 (2)}    & \multicolumn{1}{l}{0.47 (5)}   &           \\ 
318.15   & 23.91   & \multicolumn{1}{l}{3.54\0 (2)}    & \multicolumn{1}{l}{0.41 (6)}   &           \\
320      & 23.88   &                                   &                                & 0.188 (8) \\ 
328.15   & 23.58   & \multicolumn{1}{l}{4.15\0 (3)}    & \multicolumn{1}{l}{0.42 (5)}   &           \\ 
340.15   & 23.21   & \multicolumn{1}{l}{4.91\0 (3)}    & \multicolumn{1}{l}{0.29 (5)}   &           \\ 
&&&&\\[-2.0ex]
\hline
\multicolumn{5}{l}{Ethanol}\\
\hline
239      & 18.11   & \multicolumn{1}{l}{0.192 (6)}     & \multicolumn{1}{l}{3.1\0 (3)} &  \\ 
253.15   & 17.99   & \multicolumn{1}{l}{0.250 (5)}     & \multicolumn{1}{l}{2.4\0 (2)} & 0.18\0 (1) \\
263      & 17.83   & \multicolumn{1}{l}{0.363 (7)}     & \multicolumn{1}{l}{2.2\0 (3)} &  \\ 
273.15   & 17.62   & \multicolumn{1}{l}{0.493 (8)}     & \multicolumn{1}{l}{1.7\0 (2)} & 0.178 (9) \\
280      & 17.46   & \multicolumn{1}{l}{0.64\0 (1)}    & \multicolumn{1}{l}{1.4\0 (2)} &  \\ 
298.15   & 17.08   & \multicolumn{1}{l}{1.07\0 (1)}    & \multicolumn{1}{l}{1.1\0 (1)} & 0.171 (9) \\ 
308.15   & 16.88   & \multicolumn{1}{l}{1.33\0 (2)}    & \multicolumn{1}{l}{1.0\0 (2)} &  \\ 
320      & 16.66   & \multicolumn{1}{l}{1.68\0 (1)}    & \multicolumn{1}{l}{0.8\0 (1)} &  \\ 
328.15   & 16.45   & \multicolumn{1}{l}{2.01\0 (2)}    & \multicolumn{1}{l}{0.61  (8)} & 0.161 (8) \\ 
333      & 16.31   & \multicolumn{1}{l}{2.20\0 (2)}    & \multicolumn{1}{l}{0.60  (7)} &  \\
338.15   & 16.26   & \multicolumn{1}{l}{2.40\0 (2)}    & \multicolumn{1}{l}{0.47  (7)}  &  \\  
\hline \hline
\end{tabular}
\end{center}
\end{table}

\clearpage
 %Results Methanol and Ethanol
\begin{table}[ht]
\begin{center}
\caption[]{Density, self-diffusion and Maxwell-Stefan diffusion coefficient as well as shear viscosity of the mixture methanol (1) + ethanol (2) at 298.15 K and $0.1$ MPa from present molecular dynamics simulation. The number in parenthesis indicates the statistical uncertainty in the last digit.\label{resmix}}
\bigskip
\begin{tabular}{l @{\hspace{0.5cm}} c @{\hspace{0.7cm}} c @{\hspace{0.7cm}} c@{\hspace{0.7cm}} c @{\hspace{0.7cm}} c@{\hspace{0.7cm}} c@{\hspace{0.7cm}} c}
\hline \hline
&&&&\\[-2.0ex]
 \multicolumn{1}{c}{$x_{1}$\0\0}    &$\rho$ &\0\0 $ D_{1}$    &\0\0 $ D_{2}$    &\0\0$ \DMS_{12}$     & $\eta$\\ 
 &&&&\\[-2.0ex] 
$\rm mol\0\rm mol^{-1}$    &$\rm mol\0\rm l^{-1}$   &\multicolumn{1}{c}{$\rm{10^{-9}m^{-2}s^{-1}}$} &\multicolumn{1}{c}{$\rm{10^{-9}m^{-2}s^{-1}}$} &\multicolumn{1}{c}{$\rm{10^{-9}m^{-2}s^{-1}}$} &  \multicolumn{1}{c}{$10^{-3} \rm Pa\0\rm s$}\\
\hline
\0\00    & 27.02  &   &   \multicolumn{1}{l}{1.07 (1)}  & & &  &  \\ 
\0\00.143 & 26.22 & \multicolumn{1}{l}{1.37\0 (2)} & \multicolumn{1}{l}{1.12 (1)} & \multicolumn{1}{l}{1.4 (3)} & \multicolumn{1}{l}{1.17 (9)} \\ 
\0\00.305 & 25.63 & \multicolumn{1}{l}{1.52\0 (1)} & \multicolumn{1}{l}{1.25 (1)} & \multicolumn{1}{l}{1.8 (3)}& \multicolumn{1}{l}{0.76 (9)} \\ 
\0\00.500 & 25.14 & \multicolumn{1}{l}{1.73\0 (1)} & \multicolumn{1}{l}{1.44 (1)} & \multicolumn{1}{l}{1.7 (2)} & \multicolumn{1}{l}{0.62 (8)} \\ 
\0\00.570 & 24.81 & \multicolumn{1}{l}{1.82\0 (1)} & \multicolumn{1}{l}{1.53 (1)} & \multicolumn{1}{l}{1.6 (3)} & \multicolumn{1}{l}{0.64 (9)} \\ 
\0\00.754 & 24.51 & \multicolumn{1}{l}{2.07\0 (1)} & \multicolumn{1}{l}{1.72 (2)} & \multicolumn{1}{l}{2.3 (3)} & \multicolumn{1}{l}{0.70 (9)} \\ 
\0\00.796 & 23.91 & \multicolumn{1}{l}{2.14\0 (2)} & \multicolumn{1}{l}{1.82 (2)} & \multicolumn{1}{l}{2.0 (3)} & \multicolumn{1}{l}{0.59 (7)} \\ 
\0\00.890 & 23.58 & \multicolumn{1}{l}{2.28\0 (1)} & \multicolumn{1}{l}{1.90 (2)} & \multicolumn{1}{l}{2.6 (3)} & \multicolumn{1}{l}{0.60 (8)} \\ 
\0\01    & 23.21 & \multicolumn{1}{l}{4.95\0 (3)} & &  & &  &  \\ 
&&&&\\[-2.0ex]
\hline
\hline
\end{tabular}
\end{center}
\end{table}
\clearpage

 %Results VLE
\begin{table}[ht]
\begin{center}
\caption[]{Vapor-liquid equilibrium data for the mixture methanol (1) + ethanol (2) at 393.15 K and 433.15 K from present molecular dynamics simulation. The number in parenthesis indicates the statistical uncertainty in the last digit.\label{resvle}}
\bigskip
\begin{tabular}{c@{\hspace{0.4cm}}c@{\hspace{2.1cm}}c@{\hspace{0.4cm}}c@{\hspace{0.4cm}}c@{\hspace{0.4cm}}c }
\hline \hline
&&&&&\\[-2.0ex]
\multicolumn{1}{c}{$x_{1}$} &\multicolumn{1}{c}{$p$}  &$ y_{1}$ &$\rho'$ &$\rho''$  &$\Delta h^{\rm v}$ \\
\multicolumn{1}{c}{$\rm mol\0\rm mol^{-1}$}   & \multicolumn{1}{c}{\rm MPa}  & \multicolumn{1}{c}{$\rm mol\0\rm mol^{-1}$}   & \multicolumn{1}{c}{$\rm mol\0\rm l^{-1}$} &
\multicolumn{1}{c}{$\rm mol\0\rm l^{-1}$}     & \multicolumn{1}{c}{$\rm kJ\0\rm mol^{-1}$}\\
\hline
\multicolumn{6}{l}{$T$ = 393.15 K}\\
\hline
\multicolumn{1}{c}{0 \0 \0 \0}             & \multicolumn{1}{l}{0.551 (6)}         & \multicolumn{1}{l}{0}
              & \multicolumn{1}{l}{14.84 (2)}         & \multicolumn{1}{l}{0.200 (1)}        
              & \multicolumn{1}{l}{33.24 (7)}                                 \\
0.150         & \multicolumn{1}{l}{0.448 (9)}         & \multicolumn{1}{l}{0.207\0 (5)}
              & \multicolumn{1}{l}{15.62 (2)}         & \multicolumn{1}{l}{0.154 (3)}        
              & \multicolumn{1}{l}{33.86 (6)}                                 \\
0.282         & \multicolumn{1}{l}{0.482 (8)}         & \multicolumn{1}{l}{0.369\0 (2)}               
              & \multicolumn{1}{l}{16.35 (2)}         & \multicolumn{1}{l}{0.168 (3)}        
              & \multicolumn{1}{l}{33.32 (7)}  \\
0.492         & \multicolumn{1}{l}{0.538 (9)}         & \multicolumn{1}{l}{0.589\0 (2)}
              & \multicolumn{1}{l}{17.54 (3)}        & \multicolumn{1}{l}{0.185 (3)}       
              & \multicolumn{1}{l}{32.67 (8)}\\
0.678         & \multicolumn{1}{l}{0.588 (9)}         & \multicolumn{1}{l}{0.752\0 (5)}
              & \multicolumn{1}{l}{18.87 (3)}         & \multicolumn{1}{l}{0.212 (3)}        
              & \multicolumn{1}{l}{31.92 (9)}\\             
0.838         & \multicolumn{1}{l}{0.572 (9)}         & \multicolumn{1}{l}{0.884\0 (3)}
              & \multicolumn{1}{l}{19.94 (3)}         & \multicolumn{1}{l}{0.205 (3)}        
              & \multicolumn{1}{l}{31.43 (8)} \\
\multicolumn{1}{c}{1.0 \0 \0}           & \multicolumn{1}{l}{0.767 (5)}         & \multicolumn{1}{l}{1.0}
              & \multicolumn{1}{l}{21.30 (3)}         & \multicolumn{1}{l}{0.294 (1)}        
              & \multicolumn{1}{l}{29.74 (8)} \\
\hline
\multicolumn{6}{l}{$T$ = 433.15 K}\\
\hline
\multicolumn{1}{c}{0 \0 \0 \0}             & \multicolumn{1}{l}{1.32\0 (3)}         & \multicolumn{1}{l}{0}
              & \multicolumn{1}{l}{13.58 (4)}          & \multicolumn{1}{l}{0.469 (2)}        
              & \multicolumn{1}{l}{28.41 (7)}\\
0.150         & \multicolumn{1}{l}{1.29\0 (2)}         & \multicolumn{1}{l}{0.192\0 (4)}
              & \multicolumn{1}{l}{14.34 (4)}          & \multicolumn{1}{l}{0.453 (7)}        
              & \multicolumn{1}{l}{28.52 (9)}\\
0.282         & \multicolumn{1}{l}{1.38\0 (2)}         & \multicolumn{1}{l}{0.339\0 (5)}
              & \multicolumn{1}{l}{14.96 (4)}          & \multicolumn{1}{l}{0.486 (7)}        
              & \multicolumn{1}{l}{28.08 (9)}\\
0.488         & \multicolumn{1}{l}{1.48\0 (1)}         & \multicolumn{1}{l}{0.549\0 (5)}
              & \multicolumn{1}{l}{16.09 (3)}          & \multicolumn{1}{l}{0.529 (5)}        
              & \multicolumn{1}{l}{27.50 (8)}\\
0.588         & \multicolumn{1}{l}{1.54\0 (2)}         & \multicolumn{1}{l}{0.659\0 (5)}
              & \multicolumn{1}{l}{18.87 (3)}          & \multicolumn{1}{l}{0.548 (7)}        
              & \multicolumn{1}{l}{27.28 (8)}\\
0.674         & \multicolumn{1}{l}{1.58\0 (2)}         & \multicolumn{1}{l}{0.734\0 (5)}
              & \multicolumn{1}{l}{17.22 (4)}          & \multicolumn{1}{l}{0.560 (7)}        
              & \multicolumn{1}{l}{27.07 (9)}\\
0.836         & \multicolumn{1}{l}{1.64\0 (2)}         & \multicolumn{1}{l}{0.872\0 (2)}
              & \multicolumn{1}{l}{18.31 (4)}          & \multicolumn{1}{l}{0.594 (7)}        
              & \multicolumn{1}{l}{26.55 (8)}\\
0.942         & \multicolumn{1}{l}{1.72\0 (2)}         & \multicolumn{1}{l}{0.957\0 (1)}
              & \multicolumn{1}{l}{19.15 (5)}          & \multicolumn{1}{l}{0.633 (7)}        
              & \multicolumn{1}{l}{26.12 (9)}\\              
\multicolumn{1}{c}{1.0 \0 \0}           & \multicolumn{1}{l}{1.89\0 (3)}         & \multicolumn{1}{l}{1.0}
              & \multicolumn{1}{l}{19.42 (4)}          & \multicolumn{1}{l}{0.711 (2)}        
              & \multicolumn{1}{l}{26.12 (9)}\\              
&&&&&\\[-2.0ex]
\hline
\hline
\end{tabular}
\end{center}
\end{table}
\clearpage

 %Results Excess properties
\begin{table}[ht]
\begin{center}
\caption[]{Volume, excess volume, enthalpy and excess enthalpy of the mixture methanol (1) + ethanol (2) at $298.15$ K and $0.1$ MPa from present molecular dynamics simulation. The numbers in parenthesis indicate the statistical uncertainty in the last digits.\label{resexcess}}
\bigskip
\begin{tabular}{l @{\hspace{0.5cm}} c @{\hspace{0.4cm}} c @{\hspace{0.4cm}} c @{\hspace{2.8cm}} c }
\hline \hline
&&&\\[-2.0ex]
 \multicolumn{1}{c}{$x_{1}$\0\0} &\multicolumn{1}{c}{$v$} &\multicolumn{1}{c}{$v^{\rm E}$}   &\multicolumn{1}{c}{$ h$}  &\multicolumn{1}{c}{$h^{\rm E}$} \\
 &&&\\[-2.0ex] 
\multicolumn{1}{c}{$\rm mol\0\rm mol^{-1}$}    & \multicolumn{1}{c}{$\rm cm^{-3}\0\rm mol^{-1}$} &\multicolumn{1}{c}{$\rm cm^{-3}\0\rm mol^{-1}$} & \multicolumn{1}{c}{$\rm kJ\0\rm mol^{-1}$} & \multicolumn{1}{c}{$\rm J\0\rm mol^{-1}$}\\
\hline
\0\00    & \multicolumn{1}{l}{58.44\0 (5)} & &\multicolumn{1}{l}{$-43.46$\0 (4)} &  \\ 
\0\00.070 & \multicolumn{1}{l}{57.15\0 (5)} & \multicolumn{1}{l}{0.00\0 (7)} & \multicolumn{1}{l}{$-43.20$\0 (4)} & \multicolumn{1}{l}{\0 $3$\0(52)} \\ 
\0\00.200 & \multicolumn{1}{l}{54.94\0 (5)} & \multicolumn{1}{l}{0.01\0 (6)} & \multicolumn{1}{l}{$-42.75$\0 (4)} & \multicolumn{1}{l}{\0 $2$\0(49)} \\ 
\0\00.270 & \multicolumn{1}{l}{53.72\0 (5)} & \multicolumn{1}{l}{0.03\0 (6)} & \multicolumn{1}{l}{$-42.50$\0 (4)} & \multicolumn{1}{l}{$-1$\0(47)} \\ 
\0\00.350 & \multicolumn{1}{l}{52.28\0 (5)} & \multicolumn{1}{l}{0.00\0 (6)} & \multicolumn{1}{l}{$-42.22$\0 (4)} & \multicolumn{1}{l}{$-9$\0(46)} \\ 
\0\00.420 & \multicolumn{1}{l}{51.08\0 (5)} & \multicolumn{1}{l}{0.04\0 (6)} & \multicolumn{1}{l}{$-41.95$\0 (4)} & \multicolumn{1}{l}{~$20$\0(45)} \\ 
\0\00.500 & \multicolumn{1}{l}{49.65\0 (4)} & \multicolumn{1}{l}{0.02\0 (5)} & \multicolumn{1}{l}{$-41.69$\0 (4)} & \multicolumn{1}{l}{$-8$\0(44)} \\ 
\0\00.600 & \multicolumn{1}{l}{47.91\0 (4)} & \multicolumn{1}{l}{0.04\0 (5)} & \multicolumn{1}{l}{$-41.32$\0 (4)} & \multicolumn{1}{l}{\0 $5$\0(44)} \\
\0\00.697 & \multicolumn{1}{l}{46.19\0 (4)} & \multicolumn{1}{l}{0.04\0 (5)} & \multicolumn{1}{l}{$-40.96$\0 (4)} & \multicolumn{1}{l}{~$20$\0(44)} \\
\0\00.800 & \multicolumn{1}{l}{44.34\0 (4)} & \multicolumn{1}{l}{0.00\0 (5)} & \multicolumn{1}{l}{$-40.61$\0 (4)} & \multicolumn{1}{l}{\0 $1$\0(45)} \\
\0\00.905 & \multicolumn{1}{l}{42.48\0 (4)} & \multicolumn{1}{l}{0.00\0 (5)} & \multicolumn{1}{l}{$-40.24$\0 (4)} & \multicolumn{1}{l}{$-1$\0(47)} \\
\0\01    & \multicolumn{1}{l}{40.81\0 (4)} & &\multicolumn{1}{l}{$-39.90$\0 (3)}  &     \\ 
&&&\\[-2.0ex]
\hline
\hline
\end{tabular}
\end{center}
\end{table}
\clearpage

% List of figures
\listoffigures

%self-diffusion 
\clearpage

\clearpage

\clearpage

%MS Diffusion 

\clearpage

% Shear viscosity
\clearpage

\clearpage

\clearpage

% Thermal conductivity

\clearpage

\clearpage

%Velocity autocorrelation functions

\clearpage
\clearpage

\clearpage

\clearpage

\clearpage

%Shear viscosity autocorrelation functions

\clearpage

\clearpage

\clearpage
\clearpage

\clearpage
\clearpage

%\clearpage
%\begin{figure}[htb]
%\begin{center}
%\caption[Kinetic-potential term of the shear viscosity autocorrelation function and its integral for pure methanol at 0.1 MPa and $308$ K.]{} \label{figsacfkp2}
%\includegraphics[width= 150mm]{sacfkp2}
%\end{center}
%\end{figure}
%\clearpage

\clearpage
\clearpage

%radial distribution

\clearpage

% VLE

\clearpage

\clearpage

%\clearpage
%\begin{figure}[htb]
%\begin{center}
%\caption[Vapor to liquid mole fraction ratio of the mixture methanol + ethanol at $433.15$ K. Present simulation results ($\bullet$) are compared to experimental data \cite{butcher1966} ($+$). The lines represent the Peng-Robinson equation of state.]{} \label{figvlexy4}
%\includegraphics[width= 150mm]{vlexy4}
%\end{center}
%\end{figure}

%\clearpage

%Excess Properties

\clearpage

\clearpage

\clearpage
\end {document}